
\magnification=\magstep 1
\hsize=140mm 
\vsize=202mm 
\hoffset=-3mm
\voffset=-1mm
\parindent=5mm \parskip=3pt plus1pt minus.5pt
\global\newcount\secno \global\secno=0
\def\newsec#1{\global\advance\secno by1 \bigbreak\bgap
{\bbf\the\secno. #1} $\hbox{\vrule width0pt height0pt depth13pt}$
\nobreak\newline\message{(\the\secno. #1)}}

\font\ninepoint=cmr9

\font\hbf=cmbx10 scaled\magstep2
\font\bbf=cmbx10 scaled\magstep1

\font\sbf=cmbx10 scaled\magstep1

\font\sbf=cmbx9
\font\srm=cmr9

\font\sans=cmss10
\font\caps=cmcsc10
\font\nineit=cmti9
\font\ninebf=cmbx9

\def\sgap{\par\smallskip\noindent}
\def\mgap{\par\medskip\noindent}
\def\bgap{\par\vskip 8mm\noindent}  
\def\tgap{\par\vskip 20mm\noindent}
\def\newline{\hfil\break\noindent}

\def\rep{{repre\-sen\-tation}}   \def\vrep{{vacuum \rep}}
\def\qft{quantum field theory}   \def\qfl{quantum field theoretical}
\def\emor{{endo\-mor\-phism}}    \def\amor{{auto\-mor\-phism}}
    \def\imor{{iso\-mor\-phism}}
  \def\cmor{{canonical \emor}}
\def\lmor{{localized \emor}}     \def\sps{{super\-se\-lec\-tion}}
\def\vna{von Neumann algebra}    \def\cex{conditional expectation}
\def\ovw{operator valued weight} \def\pipo{Pimsner-Popa bound}
\def\jopro{Jones projection}     \def\irr{irreducible}
\def\itw{intertwiner}            \def\nos{net of subfactors}
\def\noss{nets of subfactors}

\def\aa{{\cal A}} \def\bb{{\cal B}} \def\hh{{\cal H}} \def\kk{{\cal K}}
\def\ii{{\cal J}} \def\mm{{\cal M}} \def\nn{{\cal N}} \def\oo{{\cal O}}
\def\ww{{\cal W}}
\def\sig{{\sigma}}  \def\lam{\lambda}  \def\g{\gamma}  \def\del{\delta}
\def\vphi{\varphi}  \def\eps{\varepsilon}  \def\rho{\varrho}
\def\id{{id}}
\def\comp{{\scriptstyle \circ}}  \def\inv{^{-1}}     \def\ovl{\overline}
\def\II{I\!I} \def\III{I\!I\!I}  \def\ch{_{\rm ch}}  \def\til{\tilde}
\def\ind{^{\rm ind}} \def\ext{^{\rm ext}} \def\rest{_{\rm rest}}
\def\opp{^{\rm opp}}
\def\Clebsch#1#2#3#4{{C{\textstyle{({#1#2 \atop #3})}}_#4}}
\def\vac#1{\langle\Omega,#1\Omega\rangle}
\def\qed{\vcenter{\vbox{\hrule width1.2ex height.1ex
       \hbox{\vrule width.1ex height1.0ex \kern0.97ex \vrule width.1ex}
       \hrule height.1ex}}}
\def\QED{\hfill$\qed$\mgap}
\def\semi{\setbox0=\hbox{$\times$}
     \hbox to0pt{\kern0.75\wd0 \vrule width.05ex height0.8\ht0\hss}\box0}
\def\frac#1#2{{#1\over#2}}          \def\map#1#2{\colon #1 \to #2}
\def\eins{{\mathchoice {\rm 1\mskip-4mu l} {\rm 1\mskip-4mu l}
  {\rm 1\mskip-4.5mu l} {\rm 1\mskip-5mu l}}}
\def\CC{{\mathchoice {\setbox0=\hbox{$\displaystyle\rm C$}\hbox{\hbox
  to0pt{\kern0.4\wd0\vrule height0.9\ht0\hss}\box0}}
  {\setbox0=\hbox{$\textstyle\rm C$}\hbox{\hbox
  to0pt{\kern0.4\wd0\vrule height0.9\ht0\hss}\box0}}
  {\setbox0=\hbox{$\scriptstyle\rm C$}\hbox{\hbox
  to0pt{\kern0.4\wd0\vrule height0.9\ht0\hss}\box0}}
  {\setbox0=\hbox{$\scriptscriptstyle\rm C$}\hbox{\hbox
  to0pt{\kern0.4\wd0\vrule height0.9\ht0\hss}\box0}}}}
\def\NN{{\rm I\!I\!\!N}} 
\def\QQ{{\mathchoice {\setbox0=\hbox{$\displaystyle\rm Q$}\hbox{\raise
  0.15\ht0\hbox to0pt{\kern0.4\wd0\vrule height0.8\ht0\hss}\box0}}
  {\setbox0=\hbox{$\textstyle\rm Q$}\hbox{\raise
  0.15\ht0\hbox to0pt{\kern0.4\wd0\vrule height0.8\ht0\hss}\box0}}
  {\setbox0=\hbox{$\scriptstyle\rm Q$}\hbox{\raise
  0.15\ht0\hbox to0pt{\kern0.4\wd0\vrule height0.7\ht0\hss}\box0}}
  {\setbox0=\hbox{$\scriptscriptstyle\rm Q$}\hbox{\raise
  0.15\ht0\hbox to0pt{\kern0.4\wd0\vrule height0.7\ht0\hss}\box0}}}}
\def\RR{{\rm I\!R}} 
\def\ZZ{{\mathchoice {\hbox{$\sans\textstyle Z\kern-0.43em Z$}}
  {\hbox{$\sans\textstyle Z\kern-0.43em Z$}}
  {\hbox{$\sans\scriptstyle Z\kern-0.32em Z$}}
  {\hbox{$\sans\scriptscriptstyle Z\kern-0.22em Z$}}}}   

\noindent DESY 94-205  \newline
November 1994 \hfill hep-th/9411077

\noindent (revised version Dec.\ 94)
\tgap
\centerline{\hbf Nets of subfactors}               
\vskip 4mm
\centerline{{\sl Dedicated to {\caps Bert Schroer} on the occasion of
            his 60th birthday}}
\vskip 8mm
\centerline{\caps R.\ Longo} \sgap
\centerline{Dipartimento di Matematica, Universit\`a di Roma `Tor
Vergata' ({\caps Italy})} \mgap
\centerline{\caps K.-H.\ Rehren} \sgap
\centerline{II.\ Institut f\"ur Theoretische Physik, Universit\"at
Hamburg ({\caps Germany})}
\bgap
\vskip 8mm
{\narrower \sgap
{\sbf Abstract:} 

{\srm A subtheory of a quantum field theory specifies von~Neumann
subalgebras $\aa(\oo)$ (the `observables' in the space-time region
$\oo$) of the von~Neumann algebras $\bb(\oo)$ (the `fields' localized
in $\oo$). Every local algebra being a (type $\III_1$) factor, the
inclusion $\aa(\oo) \subset \bb(\oo)$ is a subfactor. The assignment of
these local subfactors to the space-time regions is called a `net of
subfactors'. The theory of subfactors is applied to such nets.
In order to characterize the `relative position' of the subtheory,
and in particular to control the restriction and induction of
superselection sectors, the canonical endomorphism is studied. The
crucial observation is this: the canonical endomorphism of a local
subfactor extends to an endomorphism of the field net, which in turn
restricts to a localized endomorphism of the observable net. The method
allows to characterize, and reconstruct, local extensions $\bb$ of a
given theory $\aa$ in terms of the observables. Various non-trivial examples
are given. Several results go beyond the quantum field theoretical
application.}}

\baselineskip14.5pt
\def\HK{1} \def\DR{2} \def\DHR{3} \def\Tak{4} \def\BW{5} \def\Wass{6}
\def\FRS{7} \def\DL{8} \def\CD{9} \def\Hgp{10} \def\Con{11} \def\Kos{12}
\def\Jon{13} \def\PP{14} \def\Hiai{15} \def\Lon{16} \def\KL{17}
\def\Hopf{18} \def\Izu{19} \def\BDF{20} \def\Rob{21} \def\NCG{22}
\def\Popa{23} \def\RS{24} \def\BSM{25} \def\Coset{26} \def\RST{27}
\def\Ocn{28} \def\Spt{29} \def\GL{30} \def\Izpriv{31} \def\Nucl{32}
\def\NW{33}
\def\cite#1{[$#1$]}
\newsec{Introduction}
A \qft\ in the Haag-Kastler approach \cite\HK\ is given by a net of
\vna s. The net encodes the local structure of the observables of the
theory, i.e., the notion of localization going along with every
experimental operation. In this paper, we are mainly interested in a
pair of such theories of which one extends the other in a local way,
i.e., for every space-time region one has the inclusion of the
corresponding local algebras. The prototype is given by the fixpoints
under a global inner symmetry group \cite\DR, but our analysis is
focussed on the case when there is no gauge group.

The case when there is a compact gauge group is the familiar situation
in four-dimensional \qft. In \cite{\DR a}, Doplicher and Roberts have
established, by means of a harmonic analysis for operator algebras, the
existence in the field algebra of isometric `charged intertwiners' for
every \sps\ sector of the observables contained in the \vrep\ of the fields.
These intertwiners implement the corresponding DHR \emor s \cite\DHR, and
transform as irreducible tensors in the corresponding \rep\ of the gauge
group. By abstracting their specific properties within the field net, the
same authors \cite{\DR b} succeeded, by a new duality theory for compact
groups, to reconstruct the gauge group along with the field net provided
the \sps\ structure of the observables is given.

In the present article, we shall address the analogous questions when there
is no gauge group, the typical situation in less than three space-time
dimensions. Instead, we only need an appropriate \cex\ which replaces the
Haar average over the gauge group. We can again establish charged intertwiners
which, however, do not implement the sectors nor obey any generalized linear
transformation law. Yet, they admit a full-fledged harmonic analysis in the
field algebra. Furthermore, we shall characterize those \sps\ sectors of the
observables which are generated by a field algebra and show how to reconstruct
the latter in terms of observable data only (provided the index is finite).
Comparing with other approaches to generalized symmetries, we emphasize
that our theory neither assumes nor predicts that the inclusion is given by
a Hopf algebra.

In order to control the `game of restriction and induction' between
\rep s (\sps\ sectors \cite\DHR) of the two theories at hand, we apply
and generalize the theory of inclusions of \vna s, and of subfactors
in particular, to isotonous and standard nets of inclusions resp.\
subfactors (for the precise definitions, see Sect.\ 3). In doing so,
we assume the existence of a \cex\ onto the subtheory which respects
the local structure and preserves the vacuum state. This assumption is
physically motivated by the observations that $(a)$ the Haar average
over a compact gauge group which commutes with the space-time symmetry
provides a \cex\ on the gauge invariants with the claimed properties, and
that $(b)$ Takesaki's results on \cex s \cite\Tak\ imply the existence of
a \cex\ which commutes with Poincar\'e symmetries, provided the modular
conjugations w.r.t.\ wedge algebras of the larger theory in the vacuum state
generate the Poincar\'e group \cite\BW\ and preserve the subtheory. The
latter properties can be verified in interesting models \cite\Wass.

Abstracting from \qfl\ nets, we consider \noss\ in a more general
setting in Sect.\ 3. The basic result is that the existence of a
`standard' \cex\ which respects the net structure and preserves a
distinguished cyclic and separating vector state $\omega = \vac\cdot$,
guarantees the existence of simultaneous \cmor s for all subfactors in the
net which extend a given subfactor in the net (Thm.\ 3.2.). For a directed
net, this simultaneous \cmor\ extends to the $C^*$ algebra generated
by the net, where it trivializes on a large subalgebra. The latter
property amounts to a localization property in the \qfl\ context,
which is responsible for the fact that the abstract restriction takes
localized \rep s into localized \rep s (Cor.\ 3.8.). Conversely, however,
the induction in general takes localized \rep s only into half-localized
ones due to the possibility of braid statistics (Prop.\ 3.9.).

The index of the local subfactors in a net equipped with a \cex\ as
described before, is constant over the net and may be considered as
the index of the \nos. We treat the case of finite index in Sect.\ 4.
For finite index, one can set up a `harmonic analysis' in terms of
\itw s for the non-trivial sectors of the smaller net, which are
`charged fields' in the \qfl\ setting, and which are \irr\ tensor
multiplets in the case of a finite group symmetry. Even if the larger
net is local and hence the charged fields commute at space-like
distance, the corresponding \sps\ sectors may, and will in low
dimensions, have braid statistics \cite\FRS.  Locality only implies
(and actually is equivalent to) a specific eigenvalue equation (Cor.\
4.4.) for the statistics operator.

We formulate a characterization for (local) field extensions of finite
index of a given theory of observables (Thm.\ 4.8.). This
characterization is sufficiently explicit in terms of a pair of \itw s
between reducible DHR sectors of the observables, to be exploited for
a classification program. We present a family of examples for the
construction of local field extensions, which is of relevance for
two-dimensional conformal quantum field theories.

The structure of subfactors of infinite index is much less controllable,
much like infinite groups as compared with finite groups. We discuss in
Sect.\ 5 a possible compactness criterium for \noss\ in order to `stabilize'
the structure, and formulate the
expectation that this criterium be related to the split property \cite\DL\
in \qft. (The latter is known to ensure compactness of a group of gauge
\amor s.) This section, however, contains more speculations than results.
%
\newsec{Review of infinite subfactors}
In this section, we recall a number of important concepts and results
from the theory of inclusions of \vna s and the index theory for
subfactors, with special emphasis on the properly infinite case
(comprising type $\III$ factors pertinent to \qft). Although we don't give
full proofs, we take care to point out the interrelations between the
various concepts. For further details, we refer to the original literature
\cite{\CD,\Hgp,\Con,\Kos,\Jon,\PP,\Hiai,\Lon,\KL,\Hopf} or to a similar
expos\'e in \cite\Izu. We attempt to introduce a consistent notation to be
used throughout the paper. The advanced reader may skip most of this
section which, apart from a new formula for the \cmor, Prop.\ 2.9.,
contains mainly well-known material.

A \vna\ is a weakly closed algebra $M$ of operators on a Hilbert space $\hh$.
Throughout this paper, $\hh$ is assumed to be separable. A \vna\ $M$ is
a factor if its center is trivial: $M' \cap M = \CC\eins$. A factor is of
type $I$ if it admits a faithful trace with values in $\NN_0 \cup \{\infty\}$
on the projections of $M$. It is of type $\II$ if it has a trace taking
continuous values in $\RR_+ \cup \{\infty\}$ on the projections. It is of
type $\III$ if every non-zero projection $p \in M$ is the range projection of
an isometry in $w \in M$, i.e., $w^*w = \eins$ and $ww^* = p$. Clearly, this
property excludes the existence of a trace. In local \qft, the local algebras
of observables are (under very general physical assumptions \cite\BDF)
isomorphic to the unique approximately finite-dimensional type $\III_1$
factor. A factor which contains any isometry $w$ with range projection
$ww^* < \eins$ is called infinite.

Every \vna\ $M$ has a direct integral decomposition into factors. If every
factor in this decomposition is infinite, then $M$ is called properly
infinite; in other words: every central projection $p$ of $M$ strictly
dominates another projection $q < p$ such that there is a partial isometry
$w \in M$ with $w^*w = p$ and $ww^* = q$.

An inclusion of \vna s is a pair $N \subset M$ with common unit $\eins$.
It is called a subfactor if both $M$ and $N$ are factors, and it is called
properly infinite (or infinite or type $\III$) if both $M$ and $N$ are
properly infinite \vna s (or infinite or type $\III$ factors). A subfactor
is \irr\ if the relative commutant consists only of the scalars, i.e.,
$N' \cap M = \CC\eins$. If $p$ is a non-trivial projection in the relative
commutant, then the subfactor $Np \equiv N_p \subset M_p \equiv pMp$ is
called the reduced subfactor.
\mgap
{\bf 2.1.\ Conditional expectations.}\ \cite{\CD,\Hgp,\Con,\Kos}
{\bf A.} Let $N \subset M$ be a pair of \vna s with common unit. A
\cex\ $\eps \map MN$ of $M$ onto $N$ is a completely positive
normalized (i.e., unit preserving) map with the bimodule property
 $$ \eps(n^*mn) = n^*\eps(m)n \qquad (n \in N, m \in M). $$
A \cex\ is called normal if it is weakly continuous. The set of faithful
normal \cex s $\eps \map MN$ is denoted by $C(M,N)$. An arbitrary pair
$N\subset M$ may not possess any \cex\ at all, i.e., the set $C(M,N)$
may be empty. If there is any normal \cex\ for an \irr\ inclusion,
then it is unique, and it is faithful.

{\bf B.} {\it Examples:} An example of prominent interest is the
following. Let a compact group $G$ act faithfully on $M$ (`gauge
symmetry') and let $N=M^G$ be the fixpoints of this action (`gauge
invariants'). A \cex\ is given by the Haar average over the action of
the symmetry group. Dual to this example is the case $M = N\semi G$,
the crossed product by the action of a discrete group $G$ on $N$.
Here, a \cex\ is induced by the $\del$-function on $G$. Clearly, this
construction becomes singular for continuous groups. In that case,
$C(M,N)$ turns out to be empty. Provided $G$ is locally compact, one
obtains only an \ovw\ $\eta \map MN$, i.e., an unbounded (and
unnormalized) positive map with dense domain in $M_+$ satisfying the
same bimodule property as \cex s above \cite\Hgp.

{\bf C.} In general, the set of faithful normal \ovw s $\eta \map MN$ is
called $P(M,N)$. Also $P(M,N)$ may be empty. Clearly, $C(M,N) \subset
P(M,N)$.

Associated with $N\subset M$ is the inclusion of the commutants
$M' \subset N'$. Based on Haagerup's work \cite\Hgp, Connes \cite\Con\
established a canonical bijection between $P(M,N)$ and $P(N',M')$
denoted by $\eta \mapsto \eta\inv$, such that $(\eta\inv)\inv = \eta$
and $(\eta_1\comp\eta_2)\inv = \eta_2\inv\comp\eta_1\inv$ for the
composition of two \ovw s. This bijection will not, in general, relate
$C(M,N)$ with $C(N',M')$. In fact, the latter may be empty while the
former is not, or vice versa.

If $\eins$ is in the domain of $\eps\inv$, then $\eps\inv(\eins)$ lies
in the center of $M$ due to the bimodule property. Therefore it is a
scalar whenever $M$ is a factor. Declaring $\eps\inv(\eins)=\infty$
when $\eins$ is not in the domain of $\eps\inv$, Kosaki \cite\Kos\
defines the index of $\eps$ to be the number
 $$ Ind(\eps) := \eps\inv(\eins) \in [1,\infty]. $$
If the index is a finite number $\lam < \infty$, then one can normalize
$\eps\inv$ to obtain a \cex\ between the commutants
  $$\eps' := \lam\inv \cdot \eps\inv \in C(N',M'). \eqno(2.1) $$
If both $M$ and $N$ are factors and $Ind(\eps)<\infty$, then
$\eps'{}\inv = \lam(\eps\inv)\inv = \lam\eps$, and therefore $Ind(\eps')
= Ind(\eps)$.

Both our above examples yield $Ind(\eps) = |G|$ (provided the action of
the symmetry group is outer).
\mgap
{\bf 2.2.\ The Jones extension.}\ \cite\Jon\
{\bf A.} Let $N \subset M$ be an inclusion of \vna s. Choose a unit
cyclic vector $\Omega \in \hh$ such that the state $\omega = \vac\cdot$
on $M$ is faithful and invariant under a faithful \cex\ $\eps \in
C(M,N)$. Such vectors are obtained by the GNS construction from a state
$\omega = \vphi\comp\eps$ where $\vphi$ is any faithful normal state on
$N$. The projection $e_N = [N\Omega]$ on the subspace of $\hh$ generated
from $\Omega$ by $N$ lies in $N'$. The \vna\
 $$M_1 := \langle M,e_N \rangle \eqno(2.2) $$
is called the Jones extension of $M$ by $N$ (w.r.t.\ $\vphi\comp\eps$),
and $e_N$ the corresponding \jopro. As $\eps$ and $\vphi$
vary, $e_N$ and $M_1$ change only by unitary conjugation.

{\bf B.} In terms of the projection $e_N$, the map $m \mapsto \eps(m)$
can be recovered as the unique element of $N$ such that
 $$e_Nme_N = \eps(m)e_N. \eqno(2.3) $$
Eq.\ (2.3) is referred to as `$e_N$ implements $\eps$'. Obviously,
$e_N \in N' \cap M_1$. It is also the \jopro\ for the
inclusion $M_1' \subset M'$, and $N' = \langle M', e_N \rangle$ is the
Jones extension of $M'$ by its subalgebra $M_1'$. One has $N = \{e_N\}'
\cap M$.
\mgap
{\bf 2.3.\ The \pipo.}\ \cite\PP\
Let $N \subset M$ be infinite-dimensional factors. Let $\eps \in
C(M,N)$ be a faithful normal \cex. In the case of arbitrary index
$Ind(\eps) \in [1,\infty]$, the associated \ovw\ $\eps\inv \in
P(N',M')$ is characterized by the bimodule property (for $M' \subset
N'$) and the `initial value' on the \jopro\ \cite\Kos\
 $$\eps\inv(e_N) = \eins. \eqno(2.4) $$
This implies $Ind(\eps) \geq 1$ because $\eps\inv$ is positive,
and $Ind(\eps) = 1$ iff $M=N$ because $\eps\inv$ is faithful.
Note that $Me_NM$ is dense in $M_1$ and $M'e_NM'$ is dense in $N'$.

Applying this formula for $\eps' \in C(N',M')$ and $\eta = \eps'{}\inv
\in P(M,N)$ to the corresponding implementing \jopro\ $e
\equiv e_{M'} \in M$ for the subfactor $M' \subset N'$, one gets the
estimate $\eta(n^*en) = n^*\eta(e)n = n^*n \geq n^*en$ for $n \in N$,
from which one infers the operator lower bound \cite\PP\
for a \cex\ of finite index:
 $$\eps(m_+) \geq Ind(\eps)\inv m_+ \qquad (m_+ \in M_+). \eqno(2.5) $$
(Due to eq.\ (2.18) below, every element of $M_+$ is of the form $n^*en$.)
This bound is saturated by the \jopro\ $e \in M$ for
$M' \subset N'$:
 $$\eps(e) = Ind(\eps)\inv \eins. \eqno(2.6) $$
Indeed, the best lower bound (2.5) for a given \cex\ was first
introduced by Pimsner and Popa \cite\PP\ to define the index.
\mgap
{\bf 2.4.\ The minimal index.}\ \cite{\Hiai,\Lon a,\KL}
{\bf A.} Let again $N \subset M$ be infinite-dimensional factors. When
there is any \cex\ $\eps \in C(M,N)$ with finite index, then there is a
unique \cex, called the minimal \cex\ $\eps_0 \in C(M,N)$, which
minimizes the index:
 $$Ind(\eps_0) = \inf_\eps Ind(\eps) =: [M:N]. $$
$[M:N]$ is called the index of $N$ in $M$. If $\eps_0$ is minimal in
$C(M,N)$ then $\eps_0'$ is minimal in $C(N',M')$. For tensor products
$N \otimes K \subset M \otimes L$ of subfactors as well as intermediate
subfactors $N \subset M \subset L$, both the minimal \cex s and the
minimal index are multiplicative in the obvious sense. For a reducible
subfactor with $\sum p = \eins$ a partition of unity by projections
in the relative commutant $N' \cap M$, the minimal \cex s $\eps_p \in
C(M_p,N_p)$ for the reduced subfactors $Np = N_p \subset M_p = pMp$ are
given by
 $$\eps_p(m_p) = \lam_p\inv \cdot \eps_0(m_p)p \qquad (m_p \in M_p) $$
where $\lam_p = \eps_0(p) = ([M_p:N_p]/[M:N])^{1/2}$, and the square
root of the index is additive:
 $$[M:N]^\frac 12 = \sum_p [M_p:N_p]^\frac 12. \eqno(2.7) $$
For \irr\ subfactors of type $\II_1$, the minimal index coincides
with the Jones index, but not for reducible ones since in general the
trace preserving \cex\ is different from the minimal \cex.

{\bf B.} The uniqueness of the minimal \cex\ $\eps_0$ implies that
$\eps_0 \comp Ad_u = \eps_0$ for every unitary $u \in N' \cap M$.
Therefore, $\eps_0$ has the tracial property $\eps_0(xm) = \eps_0(mx)$
for $m \in M$ and $x \in N' \cap M$, and it yields a trace state on
the commutant $N' \cap M$.
\mgap
{\bf 2.5.\ The \cmor.}\ \cite\Lon\
{\bf A.} We turn now to the properly infinite case. Let $N \subset M$ be an
inclusion of properly infinite \vna s on the separable Hilbert space $\hh$.
Then there exist vectors $\Phi \in \hh$ which are cyclic and separating for
both $M$ and $N$. Choosing any such vector, let $j_N = Ad_{J_N}$ and
$j_M = Ad_{J_M}$ be the modular conjugations w.r.t.\ $\Phi$ and the
respective algebra \cite\Tak. Then
$$\g = j_Nj_M\vert_M \in End(M) \eqno(2.8)$$
maps $M$ into a subalgebra of $N$. We call $\g$ the \cmor\ associated
with the subfactor, and denote by
$$N_1 := j_Nj_M(M) \subset N, \eqno(2.9a)$$
$$M \subset M_1 := j_Mj_N(N). \eqno(2.9b)$$
the `canonical extension resp.\ restriction'. $\Phi$ is again joint
cyclic and separating for the new inclusions (2.9$a$) and (2.9$b$),
giving rise to new canonical \emor s $\rho = j_{N_1}j_N \in End(N)$ and
$\g_1 = j_Mj_{M_1} \in End(M_1)$. Since the modular conjugations
w.r.t.\ an algebra and w.r.t.\ its commutant coincide, one infers that
$J_{N_1}J_N = J_NJ_M = J_MJ_{M_1}$ and
 $$\rho = \g\vert_N \qquad {\rm and} \qquad \g = \g_1\vert_M .$$
The inclusions (2.9) are isomorphic since $N = \g_1(M_1)$ and $N_1 =
\g(M) = \g_1(M)$. We call them `dual' to the inclusion $M \subset N$,
and call $\rho = \g\vert_N$ resp.\ $\g_1$ the dual \cmor s.

{\bf B.} The canonical extension and restriction are inverse to each
other.

The dual inclusion $N_1 \subset N$ is anti-isomorphic to the inclusion
of the commutants, $M' \subset N'$, since $N_1 = j_N(M')$ and $N =
j_N(N')$. Similarly, $M_1' \subset M'$ is the anti-isomorphic image
of $N \subset M$ under $j_M$, and has the same \jopro\
$e_N \in N' \cap M_1$, since $j_M(e_N) = e_N$. If $M$ is a factor and
$N \subset M$ has finite index, one defines the dual \cex s
 $$\eps_1 = j_M\comp\eps'\comp j_M \in C(M_1,M) \qquad {\rm and}
\qquad \del = j_N\comp\eps'\comp j_N \in C(N,N_1). $$
If both $M$ and $N$, and hence also $M'$, $N'$ and $M_1$ are factors,
we conclude from 2.1.C
 $$Ind(\eps') = Ind(\eps_1) = Ind(\eps). \eqno(2.10) $$

The \cmor\ for $M' \subset N'$ is given by $\g' = j_Mj_N\vert_{N'}
\in End(N')$.

{\bf C.} {\it Example:} In a \qft\ of a Lorentz covariant local
Wightman field, let $M = \aa(\ww)$ be the algebra of observables
associated with a wedge region $\ww = \{x \in \RR^4: x^1 > |x^0|\}$,
and $N = \aa(\ww_a)$ the subalgebra associated with the wedge $\ww_a =
\{x+a : x \in \ww\}$ shifted by a vector $a$ in the positive
$x^1$-direction. The vacuum vector is cyclic and separating for both
$N$ and $M$. Since the modular conjugation for wedge regions is
(essentially) a reflection along the rim of the wedge \cite\BW,
one finds that the \cmor\ is just the translation by $2a$, and $N_1 =
\aa(\ww_{2a})$ and $M_1 = \aa(\ww_{-a})$. This example is a
self-dual subfactor.

{\bf D.} The \cmor\ $\g$ is unique up to inner conjugation with a
unitary in $N$ (depending on the choice of the joint cyclic and
separating vector $\Phi$). The canonical extension coincides with the
Jones extension w.r.t.\ $\vphi\comp\eps = \langle\Phi, \eps(\cdot)
\Phi\rangle$
 $$M_1 = j_M(N') = \langle M,e_N \rangle \eqno(2.11) $$
provided the latter is defined, i.e., provided there is a faithful
\cex\ $\eps \in C(M,N)$. (See Sect.\ 2.6.\ how to recover the latter
from $\g$.) In this case, let $e_N$ be the \jopro\ w.r.t.\ the
invariant state $\vphi\comp\eps$ where $\vphi = \langle\Phi,\cdot
\Phi\rangle$ (cf.\ Sect.\ 2.2.A). Define the isometry $v' \in N'$
by $n\Phi \mapsto n\Omega$ where $\Omega\in e_N\hh$ is the vector
representative of the invariant state $\omega = \vphi\comp\eps$ on $M$.
By construction, $J_Mv' = v'J_N$, and $v'v'^* = e_N$. Consequently
$v_1 := j_M(v')$ is an isometry in $M_1$ with the same final projection
$v_1v_1^* = j_M(e_N) = e_N$, and $J_N J_M = v'^*v_1$. This yields
another formula for the \cmor\
 $$\g(m) = Ad_{J_NJ_M}(m) = v'^*v_1 m v_1^*v' \qquad (m \in M).
                                                       \eqno(2.12) $$
\mgap
{\bf 2.6.\ Conditional expectations (resumed).}\ \cite{\Lon b}
In the properly infinite case, \cex s and \cmor s can be described in terms
of appropriate \itw s. Let $N \subset M$ be an infinite subfactor
with a faithful normal \cex\ $\eps \in C(M,N)$. Let $e_N \in N' \cap
M_1$ be the corresponding \jopro. Let $v' \in N'$ and $v_1 =
j_M(v')\in M_1$ be isometries as in Sect.\ 2.5.D. Then, $J_Mv_1 =
v_1J_{M_1}$ implies that $v_1 \map \id\g_1$ and $w = \g_1(v_1) \map \id
\rho$ are isometric \itw s in $M_1$ resp.\ in $N$ for the dual \cmor s
$\g_1 \in End(M_1)$ resp.\ $\rho = \g\vert_N \in End(N)$. In particular,
$\rho$ contains the identity $\id$ as a subsector; namely, the projection
$p=ww^*$ reduces the subfactor $\rho(N) \subset N$ to the trivial subfactor
$\rho(N)p = pNp$. The isometry $w$ induces the \cex\ by the formula
 $$\eps(m) = \eps_w(m) \equiv w^*\g(m)w. \eqno(2.13) $$
Every map of the form $\eps_w$ with $w \in N$ an isometric \itw\ $w \map
\id\rho$ is a normal \cex\ (possibly non-faithful if the subfactor is
reducible), and every \cex\ $\eps \in C(M,N)$ is of the form $\eps_w$.
\mgap
{\bf 2.7.\ Infinite subfactors of finite index.}\ \cite{\Lon b,\Hopf}
{\bf A.} A normal \cex\ $\eps \in C(M,N)$ has finite index $Ind(\eps) =
\lam$ if and only if in addition to the isometric \itw\ $w \map \id\rho$
in $N$ which $w$ induces $\eps$, i.e., $\eps = \eps_w$ as in (2.13),
there is a `dual' isometric \itw\ $v \map \id\g$ in $M$ such that
 $$w^*v = \lam^{-1/2}\eins = w^*\g(v). \eqno(2.14) $$
The isometry $v$ induces the dual \cex\ (cf.\ 2.5.B)
 $$\eps_1(m_1) = v^*\g_1(m_1)v \qquad (m_1\in M_1).  \eqno(2.15) $$
The range projection $ww^*$ saturates the \pipo\ for the dual \cex\
$\del = \g\comp\eps_1\comp\g_1\inv$, and $vv^*$ saturates the \pipo\
for $\eps$, as follows by direct computation using (2.14). The two
isometries determine each other uniquely by
 $$\lam^{-1/2} w = \eps(v) \qquad {\rm and} \qquad
   \lam^{-1/2} v = \eps_1(v_1) = \g\inv(\del(w)). \eqno(2.16) $$
Equivalently to (2.15), $w_1 = \g(v) \in N_1$ induces the dual \cex\
$\del(n) = w_1^*\rho(n)w_1$ and $w' = j_M(v) \in M'$ induces
$\eps'(n') = w'^*\g'(n')w'$.

{\bf B.} In terms of this pair of isometries, many of the previous more
general results turn into explicit algebraic relations. First, we have
already seen that the projections $vv^*$ and $ww^*$ saturate the \pipo s
for $\eps$ and $\del$.

Next, the range projection $e = vv^* \in M$ is the \jopro\ for
$N_1 \subset N$. Namely, any unit vector of the form $\Psi = v\Phi
\in e\hh$ is cyclic for $N$ provided $\Phi$ is cyclic for $M$. It
gives rise to a state $\psi = \langle \Psi, \cdot \Psi\rangle$ on $N$
which is invariant under $\del = \g(v^*)\rho(\cdot)\g(v)$, i.e.,
$\psi\comp\del = \psi$, since $v$ intertwines $v \map \id\g$. Thus,
$e = vv^*$ coincides with the projection $[N_1\Psi] \in N_1'$ on
the subspace generated from the vector $\Psi$ by $N_1$. Similarly,
$e_N = \g_1\inv(ww^*)$ recovers the \jopro\ for $N \subset M$.

Furthermore, one can compute the identity
 $$m = \lam\eps(mv^*)v = \lam v^*\eps(vm)
                                 \qquad (m \in M), \eqno(2.17) $$
which uniquely represents every $m \in M$ in the form $nv$ with $n \in
N$. Hence, we have the pointwise equality
 $$M = Nv. \eqno(2.18) $$
This yields another formula for the \cmor\ in terms of the \cex, namely
 $$\g(m) = \lam \eps(v m v^*) \qquad \quad (m \in M). \eqno(2.19) $$

{\bf C.} The subfactor $N \subset M$ is completely characterized by the
triple $(\g,v,w)$. Here $\g \in End(M)$, and the isometries $v \map
\id\g$ and $w \map \g\g^2$ are considered as \itw s in $M$ which
satisfy (2.14) and in addition
 $$ww^* = \g(w^*)w \qquad {\rm and} \qquad ww = \g(w)w. \eqno(2.20) $$
Namely, $N$ is recovered as the image of the map $\eps_w$ (cf.\ (2.13))
which turns out to be a \cex\ in $C(M,N)$.
\mgap
{\bf 2.8.\ Subfactors given by \emor s.}\ \cite{\Lon,\Rob}
{\bf A.} Let $M$ be an infinite factor and $\sig \in End(M)$ be
a unit-preserving injective \emor. Consider the subfactor
 $$N = \sig(M) \qquad (\sig \in End(M)). \eqno(2.21) $$
Whenever $M$ and $N$ are isomorphic, then $N$ can be represented as
$N = \sig(M)$ where $\sig$ is composed by an \imor\ with the injection
map $N \hookrightarrow M$.

The inner equivalence class $\sig \sim \sig'$ if $\sig' = Ad_u\comp
\sig$ with $u$ a unitary in $M$ is called a sector. If $\sig$ is \irr,
$\sig(M)'\cap M = \CC$, then an \irr\ \emor\ $\bar\sig \in End(M)$ is
called conjugate to $\sig$ if both $\sig\bar\sig$ and $\bar\sig\sig$
contain the identity as a subsector, i.e., there are isometric \itw s
$r \map \id\bar\sig\sig$ and $\bar r \map \id\sig\bar\sig$ in $M$. These
\itw s are unique up to a phase. In particular, the index of $\sig(M)
\subset M$ is finite (eq.\ (2.24) below).

If $\sig$ is reducible, then $\bar\sig$ is a conjugate if $r$ and
$\bar r$ exist such that in addition
$$\sig(r^*)\bar r = d\inv\eins = \bar\sig(\bar r^*)r \eqno(2.22) $$
holds. The conjugate is unique up to inner equivalence and depends
only on the sector of $\sig$. Clearly, $\sig$ is conjugate to
$\bar\sig$ iff $\bar\sig$ is conjugate to $\sig$.

Given the pair of isometries (2.22), the map
 $$\phi(m) = r^*\bar\sig(m)r \eqno(2.23) $$
is completely positive and normal. It satisfies
$$\phi(\sig(m_1)m\sig(m_2)) = m_1\phi(m)m_2 \qquad (m,m_i \in M) $$
and in particular $\phi\comp\sig = \id$. Such maps are called
left-inverses of $\sig$. The map $\eps_\phi := \sig\comp\phi \map M
\sig(M)$ is a \cex\ in $C(M,\sig(M))$. It satisfies the \pipo\ with
$\lam = d^2$ which is saturated by the projection $\bar r \bar r^*$,
by (2.22). Thus, the index is
$$Ind(\eps_\phi) = d^2.$$
Varying the pair $r$ and $\bar r$, one can minimize the scalar $d$.
The corresponding left-inverse is called the (unique) standard
left-inverse, \cite{\DHR,\Rob}
and the corresponding \cex\ coincides with the minimal \cex. The
minimal value of $d$ is called the `dimension' $d(\sig) \equiv d_\sig$
of $\sig$. Hence
 $$[M:\sig(M)] = d(\sig)^2. \eqno(2.24) $$

{\bf B.} If $\sig$ and $\tau$ are \emor s of $M$, then an \itw\
$x \map \sig\tau$ is an operator $x \in M$ satisfying
 $$x\sig(m) = \tau(m)x \qquad (m \in M). $$
If $\sig$ is \irr, i.e., $\sig(M)' \cap M = \CC$, then $x$ is
automatically a multiple of an isometry.

Since $M$ is infinite, it possesses orthonormal isometries
$w_i \in M$ ($i=1,\ldots n$) with complementary range projections, i.e.,
$\sum_i w_iw_i^* = \eins$. Let $\sig_i$  be \emor s of $M$. An \emor\
$\sig \simeq \bigoplus \sig_i$ is constructed by putting $\sig(m) :=
\sum_i w_i\sig_i(m)w_i^*$ which depends only up to inner equivalence on
the choice of the system of isometries. By construction, $w_i$ are \itw s
$w_i \map {\sig_i}\sig$.
Conversely, every partition of unity $\sum_i p_i = \eins$
by projections in $\sig(M)' \cap M$ gives rise to a decomposition $\sig
\simeq \bigoplus \sig_i$ by choosing isometries $w_i \in M$ with range
projections $p_i$ and defining $\sig_i(m) := w_i^*\sig(m)w_i$. Again,
$w_i \map {\sig_i}\sig$ are \itw s by construction. The reduced subfactor
$\sig(M)_{p_i} \subset M_{p_i}$ is equivalent to $\sig_i(M) \subset M$.

It follows from (2.7) that the dimension $d(\sig)$ is additive for
direct sums. It is also multiplicative for composition of \emor s and
invariant under conjugation. The multiplicativity of minimal \cex s is
equivalent to the multiplicativity $\phi_{\sig\tau} = \phi_\tau\comp
\phi_\sig$ for standard left-inverses, and the trace property of minimal
\cex s implies that \itw s between \emor s $x \map \sig\tau$ also
intertwine the corresponding standard left-inverses according to
 $$d_\sig\phi_\sig(xm) = d_\tau\phi_\tau(mx) \qquad (m \in M).
      \eqno(2.25) $$

For general subfactors, one has $[M:\g(M)] = [M:N]^2 = [N:\rho(N)]$,
and therefore
 $$[M:N] = d(\g) = d(\rho). \eqno(2.26) $$

{\bf C.} For infinite index, there is a more general notion of
conjugate \emor s. Namely, an \emor\ of $M$ is equivalent to an
$M$-bimodule, or to a `correspondence' \cite{\NCG,\Popa}.
Then the notion of a conjugate correspendence gives rise to a conjugate
\emor\ $\bar\sig$ (unique up to inner conjugation). It is found that
every \emor\ conjugate to $\sig$ is of the form \cite{\Lon b}
 $$ \bar\sig = \sig^{-1}\g $$
which is well defined since $\g(M) = N_1 \subset N = \sig(M)$.
Conversely, every \cmor\ is of the form
 $$ \g = \sig\bar\sig \eqno(2.27) $$
provided $N$ is isomorphic to $M$,
i.e., $N = \sig(M)$ for some injective \emor\ $\sig$ of $M$ (this is
always the case after tensoring both $M$ and $N$ with the same
appropriate auxiliary factor). $\bar\sig$ in (2.27) is a conjugate of
$\sig$. For finite index, both notions of conjugates coincide. The
identifications are given by $\g = \sig\bar\sig$, $\rho = \sig\comp
(\bar\sig\sig)\comp\sig\inv \in End(N)$, $w = \sig(r)$, $v = \bar r$,
$\eps_w = \eps_\phi$.

Note that the existence of an isometry $r=\sig\inv(w) \map \id\bar\sig
\sig$ is guaranteed also for infinite index by Sect.\ 2.6, provided
there is a \cex\ $\eps \in C(M,\sig(M))$. Then (2.23) still defines a
left-inverse, and $\eps_\phi = \eps_w$. The dual isometry $\bar r$ will
only exist for finite index.
\mgap
{\bf 2.9.\ Another formula for the \cmor.}\ We have encountered in this
section various formulae for the \cmor, eqs.\ (2.8), (2.12), (2.19),
(2.27), useful in different situations. We want to conclude the section
with another formula for the \cmor\ of an inclusion of properly infinite
\vna s, which we shall use in the following section.
\mgap\sl
{\bf Proposition:} Let $N \subset M$ be an inclusion of properly
infinite \vna s, $\eps \map MN$ a faithful normal \cex, and $e_N \in
N'$ the associated \jopro. The \cmor\ $\g \map MN$ is given by
 $$ \g = \Psi\inv\circ\Phi \eqno(2.28) $$
where
 $$ \Phi(m) = v_1mv_1^* \qquad (m \in M) $$
is the \imor\ of $M$ into $Ne_N$ implemented by an isometry $v_1 \in
M_1 = \langle M,e_N \rangle$ with $v_1v_1^* = e_N$, and $\Psi$ is the
\imor\ of $N$ with $Ne_N$ given by
 $$ \Psi(n) = ne_N \qquad (n \in N). $$
Every \cmor\ of $M$ into $N$ arises in this way.
\mgap\rm
{\it Proof:} Choose $v_1$ as in 2.5.D. Then $v_1M_1v_1^* \subset
e_NM_1e_N = Ne_N$, therefore $\Phi$ maps $M$ into $Ne_N$, and $\g$
given by (2.28) is well defined. By (2.12), the \cmor\ is given by
 $$ \g(m) = v'^*v_1mv_1^*v' = v'^*\Phi(m)v' $$
where $v'=j_M(v_1) \in N'$ is an isometry with final projection $e_N$.
Thus $v'$ implements the \imor\ $\Psi\inv$, proving (2.28). Every other
choice of $v_1$ amounts to a perturbation $v_1u$ with a unitary $u \in
M_1$, which changes $\g$ by conjugation with $j_M(u^*)j_Nj_M(u)$. But
$j_M(u) \in N'$, and $j_Nj_M(u) = \g_1(u)$ exhausts the unitaries of
$N$, so all \cmor s are obtained by this perturbation.
\QED
\newsec{Nets of subfactors}
In the algebraic approach of Haag and Kastler \cite\HK, a local
\qft\ is described by a net of \vna s, i.e., an assignment
 $$ \oo \mapsto \aa(\oo) $$
of \vna s $\aa(\oo)$ on a Hilbert space $\hh$ to open bounded subsets
$\oo$ of Minkowski space. This assignment is supposed to preserve
inclusions, and to be local in the sense that algebras associated
with space-like separated regions commute with each other. Locality
is often replaced by the stronger assumption of Haag duality \cite\DHR.
We shall also assume additivity, i.e., the algebra of a union of regions
is generated by the algebras of these regions.

One furthermore assumes the existence of a geometric action of the
Poincar\'e group by automorphisms and a unique cyclic vector state
$\omega = \vac\cdot$ invariant under the Poincar\'e group (the vacuum).
Consequently, the Poincar\'e group is unitarily implemented on $\hh$.
The spectrum condition postulates that the joint spectrum of the
generators of the translations lies in the forward light-cone, i.e.,
the vacuum \rep\ on $\hh$ is a positive-energy \rep. Then one has the
Reeh-Schlieder property \cite\RS\
that the vacuum vector $\Omega$ is cyclic and separating for every
local algebra $\aa(\oo)$, i.e., both $\aa(\oo)\Omega$ and $\aa(\oo)'
\Omega$ are dense subspaces of $\hh$. As a consequence, the defining
\rep\ of local observables on $\hh$ may be identified with the GNS
\rep\ associated with the state $\omega$.

Next, we consider pairs of local quantum field theories such that
one of them ($\bb$) extends the other ($\aa$) in the sense that
 $$ \aa(\oo) \subset \bb(\oo) \eqno(3.1) $$
for every region $\oo$. Here, the subspace $\hh_0\subset\hh$ generated
by $\aa(\oo)$ from the vacuum vector will be a true subspace of $\hh$,
with $\Omega$ cyclic and separating in $\hh_0$ for every $\aa(\oo)$.
(In particular, this subspace does not depend on $\oo$.) The standard
example is a theory $\bb$ of local fields on which a compact gauge
group $G$ of inner symmetries $\g_g$ acts (faithfully), and $\aa$ is
the net of local gauge invariants. If the vacuum state does not break
the symmetry, i.e., $\omega \comp \g_g = \omega$, then $\hh$ carries
the full spectrum of \rep s of the gauge group, and $\hh_0$ is the true
subspace carrying the trivial \rep\ of $G$.

The average over the gauge action with the Haar measure provides a
normal \cex\ $\eps$ for every pair of local \vna s associated with
double cone regions. One should however bear in mind that for causally
disconnected regions in general, $\eps(\bb(\oo_1\cup\oo_2))$ is not
contained in $\aa(\oo_1\cup\oo_2) = \aa(\oo_1) \vee \aa(\oo_2)$. E.g.,
if $\psi_i$ are two operators in $\bb(\oo_i)$ which are odd under a
$\ZZ_2$ symmetry, then $\psi_1\psi_2$ is even but not generated by
invariants from $\aa(\oo_i)$. For this reason, we consider only nets
over the set $\kk$ of double cones.

Abstracting from this very structureful situation, we make the following
definitions.
\mgap\sl
{\bf 3.1.\ Definition:} A net of \vna s over a partially ordered index
set $\ii$ is an assignment $\mm: i \mapsto M_i$ of \vna s on a Hilbert
space to $i \in \ii$ which preserves the order relation, i.e., $M_i
\subset M_k$ if $i \leq k$ (isotony). A net of inclusions (resp.\ of
subfactors) consists of two nets $\nn$ and $\mm$ such that for every $i
\in \ii$,
 $$ N_i \subset M_i $$
is an inclusion of \vna s (resp.\ of subfactors). We simply write $\nn
\subset \mm$. The net $\mm$ is called standard if there is a vector
$\Omega \in \hh$ which is cyclic and separating for every $M_i$. (This
property requires $M_i$ to be properly infinite.) The net
of inclusions $\nn \subset \mm$ is called standard if $\mm$ is standard
(on $\hh$) and $\nn$ is standard on a subspace $\hh_0 \subset \hh$
(with the same cyclic and separating vector $\Omega \in \hh_0$). For a
net of inclusions $\nn \subset \mm$ let $\eps$ be a consistent assignment
$i \mapsto \eps_i$ of faithful normal \cex s in $C(M_i,N_i)$. Consistency
means that $\eps_i = \eps_k\vert_{M_i}$ whenever $i \leq k$. Then we call
$\eps$ a faithful normal \cex\ of $\mm$ onto $\nn$, writing
$\eps \in C(\mm,\nn)$. $\eps$ is called standard, if it preserves the vector
state $\omega = \vac\cdot$.
\mgap\rm
On directed nets, the consistency condition formulated in Def.\ 3.1.\ is
equivalent to the {\it a priori} stronger property
 $$ \eps_j = \eps_k \qquad {\rm on} \quad M_j \cap M_k $$
(since both coincide with $\eps_m$ when $j,k \leq m$) and implies
$\eps_j(M_j \cap M_k) \subset N_j \cap N_k$.

Note that if $\nn\subset\mm$ is a net of inclusions of
infinite-dimensional \vna s, and $\eps$ a faithful normal \cex\ of
$\mm$ onto $\nn$ (in the obvious sense, i.e., for every $i \in \ii$
separately), then every faithful normal state $\vphi$ on $\nn$ gives
rise to a faithful normal state $\omega = \vphi \comp \eps$ on $\mm$
which is invariant under $\eps$. (Here, a state on a net is understood
as a consistent family of states in the obvious sense). In particular,
$\nn\subset\mm$ is standard with $(\hh,\Omega)$ the GNS construction
from $(\mm,\omega)$ (cf.\ 2.2.A). The \jopro\ $e_N$ for every single
inclusion $N_i \subset M_i$ of a standard net of inclusions does not
depend on $i \in \ii$, by the very definition of standardness, and
will be called the \jopro\ of the net of inclusions.

Therefore, the field theoretical situation described in the beginning
of this section is a standard \nos\ $\aa \subset \bb$ over the index set
$\kk = \{{\rm double \; cones}\;\oo\}$, equipped with a \cex\ $\eps \in
C(\bb,\aa)$ given by the gauge group average which is standard provided
the vacuum state $\omega = \vac\cdot$ does not break the symmetry.

Our main interest in this paper is a pair of quantum field theories as
just described, where, however, there is {\it no gauge group}, but still
a standard \cex\ $\eps \in C(\bb,\aa)$ consistent with restriction to
subregions. The condition of unbroken symmetry in the vacuum state is
replaced by the invariance property $\omega = \omega\comp\eps$. We call
this standard net of (approximately finite-dimensional type $\III_1$)
subfactors a {\it \qfl}\ \nos.

A \qfl\ net indexed by the double-cones of Minkowski space is directed.
On the contrary, `chiral' theories over the conformal light-cone $S^1$
yield non-directed nets which require a special treatment \cite{\FRS b}.

We want to note that, at least in the \qfl\ context, the invariance
property (i.e., standardness of $\eps$) is not an independent condition.
E.g., if the local subfactors are \irr, then by the uniqueness of the
\cex, $\eps$ must commute with the Poincar\'e group. Then, if the vacuum
state is the unique translation invariant state, it is automatically
invariant under $\eps$, too.

Let us also point out that the standard \cex\ of a standard \nos, if
it exists, is implemented by the \jopro, cf.\ Sect.\ 2.2.B. The attempt
to {\it define} the \cex\ by (2.3), however, will in the absence of Haag
duality for the observables yield only a \cex\ onto the dual net
$\aa^d(\oo)$ which is larger than $\aa(\oo)$. The failure of Haag duality
explicitly discussed in \cite\BSM\ is precisely of this nature.

Our first result concerns a net over an index set of two elements.
It applies to arbitrary pairs $i \leq k$ in larger index sets $\ii$,
where it is found to act as a `germ' for the subsequent results on
directed nets.
\mgap\sl
{\bf 3.2.\ Theorem:} Let $\mm = \{M \subset \til M\}$ be a standard net
of \vna s (over an index set of two elements) with joint cyclic and
separating vector $\Omega$. Let $\nn = \{N \subset \til N\}$ be another
standard net of \vna s, such that $\nn\subset\mm$ is a standard net of
inclusions with \jopro\
 $$ e_N := [N\Omega] = [\til N\Omega]. $$
Let $\eps$ be a standard \cex\ from $\mm$ onto $\nn$, i.e., $\til\eps
\in C(\til M,\til N)$ and $\eps = \til\eps\vert_M \in C(M,N)$ preserve
the vector state $\omega = \vac\cdot$ on $\til M$ resp.\ $M$. Then
every \cmor\ $\g$ of $M$ into $N$ extends to a \cmor\ $\til\g$ of $\til
M$ into $\til N$. The extension satisfies
 $$ \til\g\vert_{M' \cap \til N} = \id. \eqno(3.2) $$
\mgap\rm
{\it Proof:} For $\g$ a \cmor\ of $M$ into $N$ choose $v_1 \in M_1 =
\langle M,e_N\rangle$ as in Prop.\ 2.9. Then $\g(m)$ is the unique
element of $N$ for which
 $$ \g(m)e_N = v_1mv_1^* \qquad (m \in M, \g(m) \in N). \eqno(3.3) $$
Since $\til M_1 = \langle \til M,e_N \rangle$, the same formula
for $\til m \in \til M$ and $\til\g(\til m) \in \til N$ defines a
\cmor\ of $\til M$ into $\til N$. Moreover, $x \in M' \cap \til N$
commutes with $M$ and with $e_N$, hence with $M_1$, and
 $$ \til\g(x)e_N = v_1xv_1^* = xv_1v_1^* = xe_N \qquad (x \in M' \cap
\til N) $$
implies $\til\g(x) = x$.
\QED
If the index set is directed, i.e., for $j,k \in \ii$ there is $m \in
\ii$ with $j,k \leq m$, we associate with a net $\mm$ of \vna s the
inductive limit $C^*$ algebra $(\bigcup_{i\in\ii} M_i)^-$ and denote it
by the same symbol $\mm$. The point is that, unlike the individual
factors $M_i$, the $C^*$ algebra can have a nontrivial \rep\ theory.
In \qft, the \vrep\ is irreducible, in contrast to $\pi_0(M_i)$ having
a huge commutant. There is a class of \rep s of particular physical
interest, the DHR \sps\ sectors with finite statistics \cite\DHR\ which
are given by \lmor s with finite dimension (on the local factors)
\cite\Lon. In the sequel, we are mainly interested in the induction and
restriction of \rep s between two such $C^*$ algebras.
\mgap\sl
{\bf 3.3.\ Corollary:} Let $\nn \subset \mm$ be a directed standard
\nos\ (w.r.t.\ the vector $\Omega \in \hh$) over a directed index set
$\ii$, and $\eps \in C(\mm,\nn)$ a standard \cex. For every $i\in\ii$
there is an \emor\ $\g$ of the $C^*$ algebra $\mm$ into $\nn$ such that
$\g\vert_{M_j}$ is a \cmor\ of $M_j$ into $N_j$ whenever $i \leq j$.
Furthermore, $\g$ acts trivially on $M_i' \cap \nn$. As $i\in\ii$ varies
to $\hat\imath$, the corresponding \emor s $\g$ and $\hat\g$ are inner
equivalent by a unitary in $N_{k}$ provided $i,\hat\imath \leq k$.
\mgap\rm
{\it Proof:} According to Thm.\ 3.2., any \cmor\ $\g$ of $M_i$ into
$N_i$ extends to $M_j$ and to $M_k$ ($i \leq j,k$). The extensions
coincide on the intersection $M_j \cap M_k$ since their extension to
$M_m$ ($j,k \leq m$) is uniquely determined (given $v_1$). By
continuity $\g$ extends to $\mm$ with values in $\nn$ and is trivial
on $M_i' \cap \nn$. The last statement of Cor.\ 3.3.\ is due to the
fact, that both $\g$ and $\hat\g$ are \cmor s of $M_k$ into $N_k$. \QED
We denote by
 $$ \rho := \g\vert_\nn \in End(\nn) \eqno(3.4) $$
the restriction of $\g \in End(\mm)$ constructed in Cor.\ 3.3.
\mgap\sl
{\bf 3.4.\ Proposition:} Let $\nn\subset\mm$ be a directed standard
\nos\ with a standard \cex. Let $\g \in End(\mm)$ be associated with
$i\in\ii$ as in Cor.\ 3.3., and $\rho \in End(\nn)$ its restriction.
One has the unitary equivalences
 $$ \pi^0 \simeq \pi_0 \comp \g \qquad {\rm and} \qquad
    \pi^0\vert_\nn \simeq \pi_0 \comp \rho \eqno(3.5) $$
where $\pi^0$ is the defining \rep\ of $\mm$ on $\hh$ and $\pi_0$
the ensuing \rep\ of $\nn$ on $\hh_0 = \ovl{\nn\Omega}$.
\mgap\rm
{\it Proof:} The equivalence is established by the isometry $v_1$ in
Thm.\ 3.2.\ whose range projection is the \jopro\ $e_N =
[\nn\Omega]$. Namely, $v_1$ is a unitary map between $\hh$ and $\hh_0$
and satisfies
 $$ \g(m)e_N = v_1mv_1^* \qquad (m \in \mm). \eqno\qed $$
Thus, if $\sig$ is any \emor\ of $\mm$, then $\pi^0\comp\sig\vert_\nn
\simeq \pi_0\comp\g\sig\vert_\nn$ yields the restriction of the
corresponding \rep\ of $\mm$. We denote by
 $$ \sig\rest := \g\comp\sig\vert_\nn \eqno(3.6) $$
the mapping $End(\mm) \to End(\nn)$ corresponding to the restriction.
\mgap\sl
{\bf 3.5.\ Proposition:} Let $\nn\subset\mm$ be a directed standard \nos\
with a standard \cex\ $\eps$. Let $\vphi$ be a faithful locally normal state
on $\nn$ (i.e., normal on $N_i$) and $\vphi\comp\eps$ the induced state on
$\mm$ which is invariant under $\eps$. The corresponding GNS \rep s are
related by
 $$ (\pi_\vphi)\ind := \pi_{\vphi\comp\eps} \simeq \pi_\vphi\comp\g.
  \eqno(3.7) $$
\sgap\rm
Before we prove the proposition, we provide a lemma about the two-element
net as in Thm.\ 3.2., and an immediate corollary.
\mgap\sl
{\bf 3.6.\ Lemma.} (Notations as in Thm.\ 3.2.) Let $\rho = \g\vert_N$
and $\til\rho = \til\g\vert_{\til N}$ be the restrictions of the \cmor s;
clearly $\rho = \til\rho\vert_N$. The isometric \itw\ $w \map\id\rho$
in $N$ which induces $\eps = \eps_w$ (cf.\ (2.13)) is also an isometric
\itw\ $w \map\id\til\rho$ and induces $\til\eps$.
\mgap\rm
{\it Proof:} Since $v_1$ is an \itw\ $v_1 \map \id\g_1$
and $w = \g_1(v_1)$, we have the identities $w^*v_1 = v_1v_1^* = e_N$
and $wv_1 = v_1v_1$. The desired relations in $\til N$
 $$ \til\g(\til n)w = w \til n \qquad {\rm and} \qquad
\til\eps(\til m) = w^*\til\g(\til m)w $$
can be tested by multiplication with $e_N \in \til N'$. Indeed, we get
from (3.3)
 $$ e_N \til\g(\til n)w = v_1\til n v_1^*w = v_1\til n e_N
= v_1e_N\til n = v_1v_1^*w\til n = e_Nw\til n \qquad
(\til n \in \til N), $$
 $$ e_N\til\eps(\til m) = e_N\til m e_N = e_N \til m v_1^*w
= e_Nv_1^*\til\g(\til m)w = e_Nw^*\til\g(\til m)w \qquad
(\til m \in \til M), $$
which prove the lemma.
\QED\sl
{\bf 3.7.\ Corollary.} Let $\nn\subset\mm$ be a directed standard \nos,
$\g$ and $\rho$ as in Prop.\ 3.4. There is an isometry $w \in N_i$
which is an \itw\ $w \map\id\rho$, and which induces the \cex\ $\eps \in
C(\mm,\nn)$ by the formula
 $$ \eps(m) = w^*\g(m)w \qquad (m \in \mm). \eqno(3.8) $$
\sgap\rm
This follows immediately from the lemma. Note that when we turn to
\qfl\ nets, then the isometry $w$ is a local observable in $\aa(\oo)$.
\mgap
{\it Proof of Prop.\ 3.5.:} The unitary equivalence is given by the map
 $$ \pi_{\vphi\comp\eps}(m)\Phi_{\vphi\comp\eps} \mapsto \pi_\vphi(\g(m)w)
 \Phi_\vphi$$
(densely defined and with dense range) between the respective GNS Hil\-bert
spaces. It evidently intertwines the \rep s (3.7), and preserves the scalar
products due to (3.8).
\QED
In particular, if $\vphi = \omega\comp\sig$ is given by an \emor\
$\sig \in End(\nn)$ relative to the cyclic state $\omega = \vac\cdot$,
i.e., $\pi_\vphi = \pi_0\comp\sig$, we obtain the induced \rep\
$(\pi_0\comp\sig)\ind \sim \pi_0 \comp\sig\g$. We denote by
 $$ \sig\ind := \sig\comp\g \eqno(3.9) $$
the mapping $End(\nn) \to End(\mm)$ corresponding to the induction.

Let us now turn to \qfl\ \noss\ $\aa \subset \bb$ indexed by the
set of double-cones (or intervals), and in particular to aspects of
localization. We assume the fields to be relatively local w.r.t.\ the
observables, i.e., $\bb(\oo_1)$ commute with $\aa(\oo_2)$ when $\oo_1$ and
$\oo_2$ are at space-like distance. Then $\aa(\oo')$, the algebra of
observables in the causal complement of $\oo$, is contained in $\bb(\oo)'
\cap \aa$. The trivialization property of $\g$ in Cor.\ 3.3.\ implies
that $\rho \in End(\aa)$ acts trivially on $\aa(\oo')$. We write $\rho \in
End_\oo(\aa)$ and call $\rho$ localized in $\oo$. Moreover, $\rho$ is
transportable since the \emor\ $\hat\rho = \hat\g\vert_\aa$
(cf.\ Cor.\ 3.3.) associated with any other
double cone $\hat\oo$ is inner equivalent to $\rho$ by a unitary in
$\aa$. (Note that $\hat\oo$ need not be a Lorentz transform of $\oo$.)
Thus $\rho$ is a DHR (= localized and transportable \cite\DHR) \emor\ of
$\aa$ describing a \sps\ sector of the observables. By Prop.\ 3.4.,
this reducible sector is the restriction of the \vrep\ of $\bb$ to the
observables. We have immediately:
\mgap\sl
{\bf 3.8.\ Corollary:} Let $\aa\subset\bb$ be a directed \qfl\ \nos, such
that $\bb$ is relatively local w.r.t.\ $\aa$. Let $\g \in End(\bb)$ be
the extension of the \cmor\ of $\bb(\oo)$ into $\aa(\oo)$ as in Cor.\ 3.3.,
and $\rho \in End(\aa)$ its restriction to the observables. Then $\rho$
is localized in $\oo$ and transportable. The restriction mapping takes DHR
\emor s into DHR \emor s, i.e., $\sig\rest \in End_\oo(\aa)$ if $\sig \in
End_\oo(\bb)$, and $\sig\rest$ is transportable if $\sig$ is transportable.
\mgap\rm
The situation is less simple with induction. Note that $\sig\ind$ is not an
extension of $\sig$. Therefore, in order to formulate a statement parallel
to Cor.\ 3.8.\ for induction, we look for an \emor\ $\sig\ext$ of $\mm$
which extends a given \lmor\ $\sig$ of $\nn$ such that $\pi_0\comp\sig\ind
\simeq \pi^0\comp\sig\ext$, and check whether $\sig\ext$ can also be chosen
localized. This will not be possible in general, even when the full power of
locality is available, due to the possibility of braid statistics. The
condition of Haag duality in the following implies locality and ensures the
existence and uniqueness of statistics operators \cite\DHR. The extension
(3.10) was previously proposed by J.\ Roberts.
\mgap\sl
{\bf 3.9.\ Proposition:} Let $\aa\subset\bb$ be a directed \qfl\ \nos,
both $\aa$ and $\bb$ satisfying Haag duality, $\g \in End(\bb)$ and
$\rho \in End_\oo(\aa)$ as in Cor.\ 3.8. With every transportable \lmor\
$\sig$ of $\aa$ associate
 $$ \sig\ext := \g\inv\comp Ad_\eps\comp\sig\g \in End(\bb) \eqno(3.10) $$
where $\eps = \eps(\sig,\rho) \map {\sig\rho}{\rho\sig}$ is a unitary
statistics operator in $\aa$. Then $\sig\ext$ extends $\sig$ and
 $$ (\pi_0\comp\sig)\ind \simeq \pi_0\comp\sig\ind \equiv
\pi_0\comp\sig\g \simeq \pi^0\comp\sig\ext. \eqno(3.11) $$
$\sig\ext$ is a wedge-localized \emor\ of $\bb$. It is localized
in a double cone if and only if $\eps(\sig,\rho)\eps(\rho,\sig) =
\eins$.
\mgap\rm
{\it Proof:} We have to check that $\sig\ext$ is well-defined on
$b\in\bb(\oo)$. Increasing $\oo$, we may assume that $\rho$ and $\sig$
are also localized in $\oo$. We choose $\hat\oo$ space-like from $\oo$
and a unitary \itw\ $u \in \aa$ transporting $\sig$ to $\hat\sig$
localized in $\hat\oo$ such that $\eps(\sig,\rho) = \rho(u^*)u$. Then
on $\bb(\oo)$
$$ Ad_\eps\comp\sig\g = Ad_{\rho(u^*)}\comp\hat\sig\g =
Ad_{\rho(u^*)}\comp\g = \g\comp Ad_{u^*} \qquad {\rm on} \quad \bb(\oo),
                                                       \eqno(3.12) $$
and $\sig\ext$ is well-defined. It extends by continuity to $\bb$.
Restricted to $\aa$, we have $\sig\ext = \g\inv\comp Ad_\eps\comp
\sig\rho = \g\inv\comp\rho\sig = \sig$, hence it extends $\sig$. Now
let $\rho$ and $\sig$ be localized in $\oo_1$ and $b \in \bb(\oo_2)$ at
space-like distance from $\oo_1$. Choosing $\oo$ to contain both
$\oo_1$ and $\oo_2$, the same formula (3.12) applies, hence
$\sig\ext(b) = u^*bu$. It depends now on the relative localization of
$\oo_1$, $\oo_2$, and $\hat\oo$ whether $b$ and $u$ commute. When the
monodromy is trivial, then $\hat\oo$ can be chosen in an arbitrary
space-like direction
without changing the statistics operator $\eps$ \cite\DHR,
and hence without changing $\sig\ext$. Thus $\hat\oo$ can be chosen
such that $u$ commutes with $\bb(\oo_2)$, and $\sig\ext(b) = b$. When
the monodromy is not a scalar (a situation which arises only in $d\leq
2$ space-time dimensions \cite{\FRS a},
then $\hat\oo$ has to be chosen in a fixed space-like direction, and
the charge transporter $u$ will not commute with $b$ if $\oo_2$ lies in
the same space-like direction from $\oo$. The equivalence (3.11) is
immediate by Prop.\ 3.4., namely $\pi^0\comp\sig\ext \simeq
\pi_0\comp\g\sig\ext = Ad_{\pi_0(\eps)}\comp\pi_0\comp\sig\g$.
\QED
Note that if the monodromy $\eps(\sig,\rho)\eps(\rho,\sig)$ is a scalar,
then it is trivial, since the isometry $w \in \aa(\oo)$ intertwines
$w \map \id\rho$ (cf.\ Cor.\ 3.7.), thus $\rho$ contains the trivial
(vacuum) \emor\ $id$ as a subsector, and the monodromy must have the
eigenvalue 1 corresponding to the vacuum. Note also that in the case of
non-trivial monodromy one has two choices to define the extension
$\sig\ext$, one localized in a right wedge and the other one localized
in a left wedge, corresponding to the choice of the statistics operator.

We have established in this section the distinguished isometries $w$ and
$v_1$ associated with a \nos. They are related by $w = \g_1(v_1)$. We
have seen their respective roles for the description of induction and
restriction in terms of the \cmor\ (cf.\ the proofs of Props.\ 3.4.\ and
3.5.). The former,
as an \itw\ $w \map\id\rho$, selects the vacuum subsector contained in
$\rho$. The latter, as an operator $\hh \to \hh_0$, carries the \vrep\ of
$\aa$ contained in $\pi^0\vert_\aa$ onto the corresponding sub\rep\ of
$\pi_0\comp\rho$. In the next section, we shall introduce the `dual'
isometry $v \in \bb(\oo)$ which exists only in the case of finite index.
This operator plays the role of a `master' charged field carrying the
\sps\ charge $\rho$, i.e., all the charges of $\aa$ contained in the
\vrep\ of $\bb$.
\newsec{The case of finite index}
We study standard \noss\ equipped with a standard \cex. If the index
set is directed, then we can show that the index is constant over the
net.

We turn then to \qfl\ \noss\ of finite index. The finiteness of the
index in particular implies a finite branching of the \vrep\ of the
field algebra in restriction to the observables. The converse
implication is also true provided all \sps\ sectors of the observables
have finite statistics.

An interesting class of models is provided by current algebras over
the circle (compactified light-cone). These are generated by local
quantum fields in two-dimensional Minkowski conformal quantum field
theories with a chiral symmetry, which due to the equations of motion
depend on one chiral (light-cone) coordinate only. The nets of local
\vna s are obtained as projective \rep s of the loop group over a
given (compact, semi-simple) Lie group \cite\Wass. Then a Lie subgroup
gives rise to a subtheory, and to a \qfl\ \nos\ as described in the
previous section. After removal of the compactification point `at
infinity', this net is even directed.  The Kac characters for the
positive energy \rep s of loop groups allow to control the branching
of \rep s upon passage to subgroups, and it is well known that the
branching is finite if and only if the pair of theories is a conformal
embedding, i.e., possess the same stress-energy tensor.

Thus, the condition of finite index selects precisely the conformal
embeddings within the class of chiral current algebra models. It goes,
however, beyond the scope of this paper to treat specific models in
detail. For rigorous model analysis, we refer to \cite\Wass, and for
some instructive examples to \cite\Coset.

Let us now return to the two-element net of Thm.\ 3.2. We assume
$\nn\subset\mm$ to be a \nos. By the \pipo, the index can only decrease
when the \cex\ is restricted to a subalgebra, i.e., $Ind(\eps) \leq
Ind(\til\eps)$. The following Lemma `dual' to Lemma 3.6.\ implies the
equality of indices.
\mgap\sl
{\bf 4.1.\ Lemma.} (Notations as in Thm.\ 3.2.\ and Lemma 3.6.)
Assume in addition that $Ind(\eps) = \lam$ is finite. Then the
isometric \itw\ $v \map \id\g$ which induces the \cex\ dual to $\eps$
(cf.\ 2.7.A) is also an \itw\ $v \map\id\til\g$ and induces the \cex\
dual to $\til\eps$. In particular, $Ind(\eps) = Ind(\til\eps)$.
\mgap\rm
{\it Proof:} Since $v_1^*v = \g_1\inv(w^*\g(v))$ is a scalar due to
(2.14), we get from (3.3)
 $$ e_N\til\g(\til m)v = v_1\til mv_1^*v = v_1v_1^*v \til m =
e_N v\til m \qquad (\til m \in \til M), $$
and applying the dual \cex\ $\til\eps_1$, we conclude that $v$ is an
\itw\ $v \map \id\til\g$. By 2.7.A, $v$ induces $\til\eps_1$. By Lemma
3.6., $w \map \id\til\rho$ induces $\til\eps$. Then, since the index is
algebraically characterized by (2.14), the statements follow.
\QED\sl
{\bf 4.2.\ Corollary.} The index is constant in a directed standard
\nos\ with a standard \cex.
\mgap
{\bf 4.3.\ Corollary.} Let $\aa\subset\bb$ be a directed \qfl\ \nos,
$\g$ and $\rho$ as before. If the index $\lam = Ind(\eps)$ is finite,
then there is an isometric \itw\ $v \map \id\g$ in $\bb(\oo)$ which
satisfies the identities (2.14) with the isometric \itw\ $w \map
\id\rho$ in $\aa(\oo)$ (cf.\ Cor.\ 3.7.). Consequently,
$$ \eps(vv^*) = \lam\inv \eins \eqno(4.1) $$
and $\g$ is given on $\bb$ by the formula
$$ \g(b) = \lam\eps(vbv^*) \qquad (b \in \bb). \eqno(4.2) $$
Furthermore, every element in $\bb$ is of the form $av$ with $a \in
\aa$, namely
$$ b = \lam\eps(bv^*)v = \lam v^*\eps(vb) \qquad (b \in \bb).
\eqno(4.3) $$
\mgap\rm
{\it Proof:} The first statement is immediate from the Lemma. The
formulae follow from (2.14) by direct computation.
\QED
{}From these formulae, the trivialization of $\g$ on $\bb(\oo)' \cap
\aa$ stated in Cor.\ 3.3.\ becomes very explicit. Varying the
double-cone $\oo$ to $\hat\oo$, the unitary \itw\ $u \map \g\hat\g$
is given by $u=\lam\eps(\hat vv^*) \in \aa$ and relates the isometries
$v$ and $\hat v \in \bb(\hat\oo)$ by $\hat v = uv$. The unitary $u$ is
a localized charge transporter for $\rho$. Since by (4.3) one has
$\bb(\oo) = \aa(\oo)v$ with $v \in \bb(\oo)$, the following equality
holds for arbitrary $\hat\oo$:
$$ \bb(\hat\oo) = \aa(\hat\oo)\hat v = \aa(\hat\oo)uv \qquad {\rm where}
\qquad u = \lam\eps(\hat vv^*). \eqno(4.4) $$
For $\sig$ a \lmor\ of the observables, we call $\psi \in \bb$ a
charged \itw\ if it satisfies the commutation relation
$$ \psi a = \sig(a) \psi. \eqno (4.5) $$
Such operators interpolate the \rep\ $\pi_0\comp\sig$ with the \vrep\
$\pi_0$ of $\aa$ within the \vrep\ $\pi^0$ of $\bb$. Charged \itw s do
not exist for every \lmor\ of $\aa$ (see below). For \irr\ subfactors,
charged \itw s are multiples of isometries since $\psi_s^*\psi_s$
commute with $\aa$ and hence are scalars.
\mgap\sl
{\bf 4.4.\ Corollary.} The isometry $v$ is a charged \itw\ for the
\lmor\ $\rho$. The isometry $w_1 = \g(v)$ is an \itw\ $w_1 \map
\rho\rho^2$ in $\aa(\oo)$. Let $\eps_\rho = \eps(\rho,\rho)$ be the
statistics operator for the \lmor\ $\rho$. Then the following
identities hold.
$$ \eps_\rho v v = v v \qquad {\rm and} \qquad \eps_\rho w_1 = w_1.
\eqno(4.6) $$
\sgap\rm
{\it Proof:} The stated intertwining properties of $v$ and $w_1$ are
obvious. The statistics operator is of the form $\rho(u^*)u$ where
$u \map \rho\hat\rho$ is a unitary charge transporter as in (4.4)
with $\hat\oo$
space-like separated from $\oo$. Then $\hat v = uv$ and $v$ commute,
$uvv = vuv = \rho(u)vv$, implying the first identity. Furthermore,
since $vv = \g(v)v = w_1v$, we have $\eps_\rho w_1v = w_1v$ which
implies the second identity by right multiplication with $v^*$ and
application of the \cex\ $\eps$.
\QED
Since the index is finite, the dimension of $\rho$ is finite, and
$\rho \in End(\aa)$ can be only finitely reducible. Let
$$ \rho \simeq \bigoplus_s N_s \rho_s $$
be the sector decomposition of $\rho$ with $\rho_s$ \irr\ and
inequivalent, and $N_s$ finite multiplicities. We shall assume from
now on that the local subfactors $\aa(\oo)\subset\bb(\oo)$ are \irr,
thus the \cex\ $\eps$ is unique and minimizes the index (cf.\ Sect.\
2.5.). Then the multiplicity of $id\prec\rho$ is 1, and the index is
given (cf.\ Sect.\ 2.8.B) by:
$$ \lam = [\bb:\aa] = \sum_s N_s d(\rho_s). $$
\sgap\sl
{\bf 4.5.\ Lemma:} There is an anti-\imor\ between the linear space of
isometries $w_s \map {\rho_s}\rho$ and the linear space of charged \itw s
$\psi_s$ for $\rho_s$, given by
$$ \psi_s = w_s^*v \qquad \Leftrightarrow
                   \qquad w_s = \lam\eps(v\psi_s^*). \eqno(4.7) $$
\sgap\rm
{\it Proof:} By direct computation.
\QED
Except for a common scale, the map (4.7) takes isometries into isometries.
In fact, we may compute $\psi_s^*\psi_s = \g(\psi_s^*\psi_s) =
\g(v^*w_sw_s^*v) = \del(p_s) = d_s/\lam$, since $p_s = w_sw_s^*$ is a
minimal projection in the relative commutant $\rho(\aa)'\cap\aa$, and
we can apply formulae from Sects.\ 2.4.A, 2.7.A, and 2.8.B. The
following conclusion is immediate.
\mgap\sl
{\bf 4.6.\ Corollary:} The multiplicity $N_s$ of a subsector $\rho_s
\in End(N)$ in the dual \cmor\ $\rho$ of an \irr\ subfactor $N\subset
M$ with finite index equals the linear dimension of the space of charged
\itw s for $\rho_s$ in $M$. It is bounded by the dimension $d(\rho_s)$,
i.e., $N_s \leq d(\rho_s)$. (Analogous statements hold for the \cmor\
$\g$.)
\mgap\rm
{\it Proof:} The first part is due to the Lemma 4.5. The \cmor\ $\g$ maps
the space of charged \itw s $\psi_s$ for $\rho_s$ into the
space of \itw s $\g(\psi_s) \map \rho\rho\rho_s$ in $\aa$. Thus
$\rho\rho_s$ contains $N_s$ orthogonal copies of $\rho$, and
$d(\rho\rho_s) \geq N_sd(\rho)$. This proves the claim.
\QED
If $v \in \bb(\oo)$ and $w_s \in \bb(\oo)$, i.e., $\rho$ and $\rho_s$
are localized in $\oo$, then the charged \itw s are elements of
$\bb(\oo)$. If $u_s$ is a unitary charge transporter to $\hat\rho_s$
localized in $\aa(\hat\oo)$ then $\hat\psi_s = u_s\psi$ are charged
\itw s for $\hat\rho_s$ in $\bb(\hat\oo)$. From (4.6) one concludes the
commutation relation
$$ \eps(\rho_s,\rho_t)\psi_s\psi_t = \psi_t\psi_s $$
which holds in spite of the fact that the monodromy operator
$\eps(\rho_t,\rho_s)\eps(\rho_s,\rho_t)$ needs not be trivial. In fact,
the range projections of charged \itw s do not exhaust the Hilbert
space in general. It was proven that the joint range projections $e_s =
\sum_i \psi_s^i\psi_s^{i*}$ (in an orthonormal basis) are all unity if
and only if $N_s = d(\rho_s)$ for all subsectors $\rho_s$ of $\rho$
\cite\Coset, and in turn if and only if the local subfactors are
fixpoint subfactors under the action of a Hopf $C^*$ algebra \cite\Hopf.
In this special case, the charged \itw s transform linearly as \irr\ tensor
multiplets under the action of the Hopf algebra and implement the \emor\
\cite\DR:
$$ \rho_s(a) = \sum_i \psi_s^i a \psi_s^{i*}. \eqno(4.8) $$
In the general case, however, and in particular when the dimensions
$d(\rho_s)$ assume non-integer values, there is neither a transformation law
nor do the charged intertwiners implement the sectors (cf.\ Sect.\ 5).

By construction of the charged \itw s, eq.\ (4.7), we have the expansion
$$ v = \sum_{s,i} w_s^i\psi_s^i, \eqno(4.9) $$
where $w_s^i$ are an orthonormal basis of intertwiners $w_s \map
{\rho_s}\rho$, $i = 1,\ldots N_s$. Inserting this into (4.3), one
obtains the expansion of arbitrary elements of $\bb$
$$ b = \lam\sum_{s,i} \eps(b\psi_s^{i*})\psi_s^i \qquad (b \in \bb).
                                                 \eqno(4.10) $$
One may consider this expansion as a generalized `harmonic analysis'
which decomposes an arbitrary operator into `\irr\ tensor multiplets'
with invariant (observable) coefficients. Note that no transformation
law has to be specified to make sense of this interpretation, nor will
it in general exist.

Let us apply the expansion (4.10) to products of charged \itw s. One
obtains coefficients in $\aa$ of the form $T =
\eps(\psi_s\psi_t\psi_u^*)$ which are \itw s $\rho_u\to \rho_s\rho_t$
and can, in turn, be expanded into an orthonormal basis of such \itw s
with complex `Clebsch-Gordan' coefficients. We summarize:
\mgap\sl
{\bf 4.7.\ Proposition.} The charged \itw s satisfy operator product
expansions
$$ \psi_s^i\psi_t^j = \lam\sum_{u,e,k} \Clebsch ijke \;
T_e \psi_u^k \qquad\qquad (\Clebsch ijke \in \CC) \eqno(4.11) $$
where $T_e$ are a basis of observable \itw s $T_e \map {\rho_u}
\rho_s\rho_t$, and only charges $\rho_u$ contribute which are contained
in $\rho$. The complex `Clebsch-Gordan' coefficients are given by
 $$ \Clebsch ijke \eins = T_e^*\eps(\psi_s^i\psi_t^j\psi_u^{k*}). $$
\mgap\rm
These operator product expansions were used in \cite\RST\
to compute correlation functions of charged fields as expansions into
`partial waves' which are determined by the subtheory $\aa$, while only
the numerical Clebsch-Gordan expansion coefficients refer to the
theory $\bb$ and are in fact completely determined by the structure of
a single local subfactor $\aa(\oo)\subset\bb(\oo)$.

Let us next compute the isometry $w_1 = \g(v)$, Cor.\ 4.4. We have
$\g(v) = \lam\eps(vvv^*)$, and with the previous expansion (4.9) of $v$,
we obtain
\mgap\sl
{\bf 4.8.\ Corollary.} \hfill
$ \displaystyle w_1 \equiv \g(v) = \lam\sum_{stu,e,ijk} \Clebsch ijke \;
\rho(w_s^i)w_t^j T_e w_u^{k*} $ \hfill {\rm (4.12)} \newline
where each term is an \itw\ $\til T \map \rho\rho^2$. In particular,
knowledge of $w_1$ determines the Clebsch-Gordan coefficients in
Prop.\ 4.7.
\mgap\rm
The significance of the isometry $w_1$ is the following. As in 2.7.C,
the triple $(\rho,w,w_1)$ uniquely characterizes the subfactor
$N_1 \subset N$, and therefore by the canonical extension also
$N \subset M$, provided $w \map \id\rho$ and $w_1 \map \rho\rho^2$
are isometries satisfying the dual analogue of (2.14), (2.20), i.e.,
$$ w_1^*w = \lam^{-1/2}\eins = w_1^*\rho(w) \eqno(4.13a) $$
$$ w_1w_1^* = \rho(w_1^*)w_1 \qquad
{\rm and} \qquad w_1w_1 = \rho(w_1)w_1. \eqno(4.13b). $$
In our present setting, the system (4.13) refers to the net of
observables only, so the field nets which extend a given observable net
can be characterized, and possibly classified, as solutions to (4.13)
where for every candidate \emor\ $\rho$ of finite dimension, the
isometries $w$ and $w_1$ are elements of finite-dimensional spaces of
\itw s. Therefore, the system (4.13) involves only finitely many
complex coefficients as unknowns.

Actually, one has to reconstruct the entire net $\bb$ rather than only
$\aa(\oo)\subset\bb(\oo)$. The requirement that $\bb$ is a local net
imposes an additional condition. We have
\mgap\sl
{\bf 4.9.\ Theorem.} Let $\aa$ be a local net of observables, $\rho$ a
localized and transportable \emor\ of $\aa$ with finite statistics,
which contains $id\prec\rho$ with multiplicity one, i.e., there is a
unique (up to a phase) isometry $w \map \id\rho$. Then $\rho$ is the
(dual) \cmor\ associated with a standard net $\aa \subset \bb$ of
\irr\ subfactors $\aa(\oo)\subset\bb(\oo)$ equipped with a standard
\cex\ $\eps$, if and only if there is an isometry $w_1 \map \rho\rho^2$
which solves the system (4.13). The index of this \nos\ equals $\lam =
d(\rho)$. The net $\bb$ is relatively local w.r.t.\ $\aa$, and it is
itself local if and only if, in addition, $\eps_\rho w_1 = w_1$ holds.
\mgap\rm
{\it Proof:} The proof is constructive, reversing the previous
discussions which established the triple $(\rho,w,w_1)$ from the \nos.
Let $\rho$ be localized in $\oo$. Put $N := \aa(\oo)$. One first
constructs the dual subfactor $N_1 \subset N$ and recovers $M =:
\bb(\oo)$ as described after (4.13). Then $\bb(\oo)=\aa(\oo)v$ with
an isometry $v \in \bb(\oo)$ which is a charged \itw\ $v \map \id\g$ for
$\g(av):=\rho(a)w_1$ on $\bb$. For generic double cones $\hat\oo$ one
chooses a unitary charge transporter $u \map \rho\hat\rho$ where
$\hat\rho$ is localized in $\hat\oo$, and defines $\bb(\hat\oo):=
\aa(\hat\oo)uv$. It is easy to verify that $\bb$ is an isotonous net
which extends $\aa$, and that $\hat\g$ defined on $\bb(\hat\oo)$ by
$\hat\g(\hat a uv):= \hat\rho(\hat a) u\rho(u)w_1u^*$ is a \cmor\ of
$\bb(\hat\oo)$ into $\aa(\hat\oo)$ which extends $\hat\rho$. Defining
$\eps$ by $\eps = w^*\g(\cdot)w$ on $\bb$, and extending the vacuum
state $\omega$ on $\aa$ to $\omega\comp\eps$ on $\bb$, one obtains a
standard net with a standard \cex. Since $uva = \hat\rho(a)uv$ and
$\hat\rho$ is localized in $\hat\oo$, fields $uv \in \bb(\hat\oo)$
commute with observables $a$ which are localized at space-like distance
from $\hat\oo$. Finally, the locality of the net
$\bb$ requires that $u_1v$ and $u_2v$ commute if $\hat\oo_1$ and
$\hat\oo_2$ are at space-like distance. An argument similar to the
one leading to (4.6) shows that this is equivalent to $\eps_\rho w_1 =
w_1$.
\QED
{\it Example:} Consider a model with an irreducible \sps\ sector
$\sig$ such that $\sig^2 \simeq \id \oplus \sig$. (There are known
models where $\sig$ is the only nontrivial sector.) Then $\rho =
\sig^2$ satisfies all the requirements of Thm.\ 4.9., since $\sig$
is its own conjugate, and $w=r$ and $w_1 = \sig(r)$ where $r$ is the
isometric intertwiner $r \map \id\sig^2$. Thus the theorem provides a
field algebra which accounts for the charged sector $\sig$ of the
observables. The index of the extension is $d^2 = d+1$ where $d =
d(\sig) = 2\cos\frac\pi 5$ is the statistical dimension of the sector
$\sig$. To be more specific:
On the Hilbert space $\hh = \hh_0\oplus\hh_\sig$ carrying the \rep\
$\pi_0 \oplus \pi_0\comp\sig$, the observables $a \in \aa$ are embedded
into the field algebra $\bb$ by
$a \hookrightarrow \pmatrix{a & 0 \cr 0 & \sig(a) \cr}$.
In particular, we have $w \in \bb(\oo)$ and $w_1 \in \bb(\oo)$ if $\sig$
is localized in $\oo$. The field algebra is then generated by the
observables and one further element $v \in \bb(\oo)$, the charged
intertwiner for $\rho$, cf.\ Cor.\ 4.4. It is explicitly given by
 $$ v = d^{-1/2} \pmatrix{d^{-1/2} r & d^{-1/2} t \cr tr & tt \cr}  $$
where $t$ is an isometric intertwiner $t \map \sig\sig^2$, and the \emor\
$\sig$ acts by
 $$ \sig(r) = d\inv r + d^{-1/2} tt \qquad {\rm and} \qquad
 \sig(t) = trr^* + d^{-1/2}rt^* - d\inv ttt^*. $$
The \cmor\ is defined on the observables by $\g(a) = \rho(a)$, and on
$v$ by $\g(v) = w_1$. The reader is invited to check all the stated
relations in terms of these informations, and to compute the charged
intertwiner for the sector $\sig$, $\psi_\sig = d^{-3/2}
\pmatrix{0 & d^{1/2}\eins \cr dr & -t \cr}$ as well as the operator
product expansions $r^*\psi_\sig^2 \propto \eins$ and $t^* \psi_\sig^2
\propto -\psi_\sig$ (cf.\ eq.\ (4.11)). Note that the condition
$\eps_\rho w_1 = w_1$ cannot be satisfied with the \sps\ structure at
hand, hence $\bb$ is not a {\it local} net.

We give now an example for the construction of a {\it local} extension
of a given local net, by solving the system (4.13), (4.6). It is
based on the following general result.
\mgap\sl
{\bf 4.10.\ Proposition.} Let $M$ be a type $\III$ factor, and $\Delta$
a finite set of inequivalent \irr\ sectors of $M$ with representatives
$\rho_s \in End(M)$ with finite dimensions $d_s = d(\rho_s)$, such that
along with every sector also the conjugate sector is in $\Delta$, and
every product of sectors in $\Delta$ is equivalent to a direct sum of
sectors in $\Delta$. Let $j$ be the natural anti-\imor\ $m \mapsto m^*$
of $M$ with the opposite algebra $M\opp$. Then the \emor\ of $A := M
\otimes M\opp$ given by
$$ \rho \simeq \bigoplus_{s \in \Delta} \rho_s
\otimes (j\comp\rho_s\comp j\inv) \eqno(4.14) $$
is the \cmor\ of $A$ into a subfactor $A_1 \subset A$ with index
$\lam = \sum_s d_s^2$.
\mgap\rm
{\it Remarks:} The conclusion remains true if $j$ is replaced by any other
anti-\imor\ between $M$ and $M\opp$ since $\rho$ (as a sector) varies only
by conjugation with $\id \otimes \alpha \in Aut(A)$ where $\alpha$ is an
\amor\ of $M\opp$. The analogous statement holds for an \emor\ $\rho$
of $M \otimes M'$, exploiting the (natural) \imor\ $j_M \comp j\inv$ between
$M\opp$ and $M'$ (replacing $j$ by $j_M$ in the proposition). If $M$ happens
to be anti-isomorphic with itself (like the approximately finite-dimensional
type $\III_1$ factor), then the analogous statement also remains true for
$\rho$ as an \emor\ of $M \otimes M$, replacing $j$ by an anti-\amor\ of $M$.
We shall consider this case in a \qfl\ application below.

{\it Proof of Prop.\ 4.10.:} According to 2.7.C, we have to establish a pair
of isometric \itw s $w \map \id\rho$ and $w_1 \map \rho\rho^2$ in $A$ which
solve the system (2.14), (2.20) for $A_1 \subset A$, i.e., (4.13).

Let $\sig \simeq \bigoplus_{s \in \Delta} \rho_s \in End(M)$ with isometric
\itw s $w_s \map {\rho_s}\sig$, in particular $w_0 \map \id\sig$. Put
$\rho_s\opp := j\comp\rho_s\comp j\inv$ and $\sig\opp := j\comp\sig\comp
j\inv$, hence $j(w_s) \map {\rho_s\opp} \sig\opp$, and put $\hat\rho :=
\sig\otimes\sig\opp \in End(A)$. Then we have projections $p_s = w_sw_s^*$
and $j(p_s)$ in the relative commutants of $\sig$ and $\sig\opp$, and $\rho$
given by (4.14) corresponds to the projection $p = \sum_s p_s \otimes j(p_s)
\in \hat\rho(A)' \cap A$ in the relative commutant of $\hat\rho$.

It is now sufficient to prove the existence of an isometric \itw\
$\hat w \map \id\hat\rho$ and a partially isometric \itw\ $\hat w_1 \map
{\hat\rho}\hat\rho^2$ with $\hat w_1^*\hat w_1 = p$ which satisfy
$$ (a) \quad p\hat w = \hat w \qquad\qquad (a_1)
\quad p\hat w_1 = \hat\rho(p)\hat w_1 = \hat w_1p = \hat w_1 $$
$$ (b) \qquad \hat w^*\hat w_1 = \lam^{-1/2}p = \hat\rho(\hat w^*)
\hat w_1 $$
$$ (c) \quad \hat w_1\hat w_1^* = \hat\rho(\hat w_1^*)\hat w_1 \qquad
\qquad (d) \quad \hat w_1\hat w_1 = \hat\rho(\hat w_1)\hat w_1 $$
since the desired identities (4.13) then follow for $w:=u^*\hat w$
and $w_1:=\rho(u^*)u^*\hat w_1u$ where $u \map \rho\hat\rho$ is
an isometry in $A$ such that $uu^* = p$. The derivation of (4.13)
from $(a) - (d)$ is straightforward.

Let us now construct $\hat w$ and $\hat w_1$. The former, $\hat w := w_0
\otimes j(w_0)$ with $w_0 \map \id\sig$ as before, is uniquely
determined up to a phase. Clearly, $\hat w \map \id\hat\rho$ is an
isometry satisfying $(a)$. For the latter, choose a system of
orthonormal bases of isometric \itw s $T_e \map {\rho_u}{\rho_t\rho_s}$
for every triple $s,t,u\in\Delta$, which are `lifted' to
$\til T_e := \sig(w_s)w_tT_eW_u^* \map \sig\sig^2$. Clearly,
$$ \hat w_1 := \lam^{-1/2}\sum_e\sqrt{d(e)} \til T_e \otimes
j(\til T_e) \qquad\qquad {\rm with} \qquad d(e) = d_sd_t/d_u $$
is an \itw\ $\hat w_1 \map {\hat\rho}\hat\rho^2$ which satisfies
$(a_1)$. It does not depend on the choice of the bases since a unitary
basis change is absorbed due to the anti-linearity of $j$.
Since $\til T_e^*\til T_{e'} = \del_{ee'}p_u$, we have
$$ \lam \hat w_1^*\hat w_1 = \sum_e \frac{d_sd_t}{d_u} p_u
\otimes j(p_u) = \sum_u(\sum_{st}N_{st}^u \frac{d_sd_t}{d_u})\;
p_u \otimes j(p_u), $$
where $N_{st}^u$ are the multiplicities of $\rho_u$ in $\rho_t\rho_s$.
Using Frobenius reciprocity, $N_{st}^u = N_{\bar s u}^t$, and the
additivity of dimensions 2.8.B, the sum over $t$ can be evaluated:
$\sum_t N_{\bar s u}^t d_t = d_sd_u$. Then the sum over $s$ gives $\lam
= \sum_s d_s^2$, and the sum over $u$ finally gives the projection $p =
\sum_u p_u \otimes j(p_u)$. We conclude that $\hat w_1^*\hat w_1 = p$ as
required.

Let us now turn to $(b)$. Due to orthogonality of the $w_s$, only $T_e$
with $\rho_t = \id$ resp.\ $\rho_s = \id$ will contribute to $\hat w^*
\hat w_1$ resp.\ $\hat\rho(\hat w^*)\hat w_1$. Without restriction we
may assume that these $T_e = \eins$, hence $\til T_e = w_0 p_s$ resp.\
$\til T_e = \sig(w_0) p_t$. Since furthermore $d(e) = 1$ for these
terms, one immediately obtains $(b)$.

Comparing the two sides of $(c)$ when the definition of $\hat w_1$
is inserted, one has to consider the change of \itw\ bases
$$ (d(f)d(g))^{1/4} \til T_f \til T_g^* \mapsto (d(e)d(h))^{1/4}
\sig(\til T_e^*) \til T_h $$
where $T_e \map {\rho_u}\rho_t\rho_s$, $T_f \map {\rho_w}\rho_r\rho_u$,
$T_g \map {\rho_w}\rho_v\rho_s$ and $T_h \map {\rho_v}\rho_r\rho_t$.
Removing the isometries $w_a$ and a common factor
$(d_vd_s/d_ud_r)^{1/4}$ from both sides, one has to consider the change
of bases $\sqrt{d(f)}T_fT_g^* \mapsto \sqrt{d(h)}\rho_r(T_e^*)T_h$
which both span the spaces of \itw s $T \map
{\rho_v\rho_s}\rho_r\rho_u$. In the inner product $\langle T,T' \rangle
= \phi_u\phi_r(T'T^*)$, both sets of \itw s are orthonormal. Namely, by
(2.25), $d(h) \phi_t\phi_r(T_{h'}T_h^*) = \phi_v(T_h^*T_{h'}) =
\del_{hh'}$, and since $\phi_r(T_{h'}T_h^*) \in \rho_t(M)'\cap M$ is a
scalar, one also has $d(h)\phi_r(T_{h'}T_h^*) = \del_{hh'}$. Similarly,
$d(f)\phi_u\phi_r(T_{f'}T_f^*) = \del_{ff'}$. With these formulae, the
orthonormality of the above bases is straightforward. Hence, for every
fixed set $v,s,r,u$, the change of basis is a unitary one, and the
transition coefficients for the two tensor factors cancel each other
due to the anti-linearity of $j$.

Similarly, for $(d)$ one has to consider the change of bases
$$ (d(g)d(h))^{1/4}T_hT_g \mapsto (d(e)d(f))^{1/4}\rho_r(T_e)T_f $$
for the spaces of \itw s $T \map {\rho_w}{\rho_r\rho_t\rho_s}$.
Observing that $d(e)d(f) = d(g)d(h) = d_rd_sd_t/d_w$ are constant in
these spaces, and both $T_hT_g$ and $\rho_r(T_e)T_f$ are orthonormal
bases in the inner product $\langle T,T' \rangle = T^*T'$, the same
argument as before applies to prove $(d)$.
\QED
We observe a structural similarity of (4.14) with the \cmor\ of
Ocneanu's `asymptotic inclusion' $M \vee (M'\cap M_\infty) \subset
M_\infty$ \cite\Ocn.
Indeed, we conjecture that the asymptotic subfactor is covered by
our proposition, where $\Delta$ is the set of \irr\ subsectors of
$\g^n$ ($n \in \NN$) of a finite depth subfactor. However, it is easy
to find examples for (4.14) which do not describe an asymptotic subfactor.

Of physical interest is the situation when $\Delta$ is a set of \sps\
sectors of a chiral \qft\ $\aa\ch$ (given as a directed net over the
set $\ii\subset\RR$ of intervals), e.g., all DHR sectors in a
`rational' such theory \cite{\Spt,\RST}. One chooses an appropriate
modular (CPT) conjugation $j$ which maps the algebra of an interval
onto the algebra of a reflected interval and takes DHR \emor s into
conjugate DHR sectors \cite{\BW,\GL}. $j$ is an anti-\amor\ of $M =
\aa(\ii)$ if the interval $I$ is symmetric under the reflection. Thus,
a variant of Prop.\ 4.10.\ (cf.\ the remarks following the proposition)
applies.

In this situation, the calculation leading to Prop.\ 4.10.\ literally
applies to the DHR \emor s $\rho_s$ of $\aa$ localized in the interval
$I$, and therefore provides a triple $(\rho,w,w_1)$ as in Thm.\ 4.9.\ for
the tensor product of two chiral theories $\aa = \aa\ch \otimes \aa\ch$.
This theory is naturally given as a net over the double cones $\oo$
which are Cartesian products of two chiral light-cone intervals. The
triple therefore gives rise, by Thm.\ 4.9., to a \qfl\ \nos\ $\aa
\subset \bb$ where $\bb$ is a two-dimensional Minkowski \qft.
Note that in the construction (4.14), $j \comp \rho_s \comp j\inv =
\bar\rho_s$ yields the conjugate sectors.

A similar argument as in the proof of Prop.\ 4.10.\ involving unitary
basis changes $T_e \mapsto \eps(\rho_t,\rho_s)T_e$ shows that $\hat w_1$
is invariant under the statistics operator $\eps_{\hat\rho} = \eps_\sig
\otimes j(\eps_\sig)$. The $CPT$ symmetry shows that $j(\eps_\sig) =
\eps_{\bar\sig}^*$, and therefore $\eps_{\hat\rho}$ involves charge
transporters in opposite light-like directions on the two chiral
light-cones, i.e., in space-like directions in two dimensions. We
conclude that $\eps_\rho := \rho(u^*)u^*\eps_{\hat\rho}u\rho(u)$
(with $u \map \rho\hat\rho$ as in the proof of Prop.\ 4.10.) is the
two-dimensional statistics operator, and satisfies $\eps_\rho w_1 =
w_1$. Therefore the net $\bb$ is a local extension of $\aa$.
%
\newsec{Comments on `compact' subfactors}
In the case of infinite index, much less can be done beyond the general
analysis in Sect.\ 3 without further assumptions. In this article, we
want to restrict ourselves to some comments only, and hope to return to
the issue on another occasion. Let us concentrate on \qfl\ \noss.

It is well known that in four-dimensional theories \cite{\DR b}
the algebra of observables can be embedded as a \nos\ into a local or
graded local (fermionic) field algebra with a compact gauge group such
that the observables are the fixpoints under the gauge symmetry. The
(dual) canonical \emor\ $\rho$ in this case is the direct sum over all
DHR \sps\ sectors of the observables with multiplicities equal to the
statistical dimension, and in turn equal to the associated \rep\ of the
gauge group. The gauge group may well be an infinite compact group in
which case $\rho$ is clearly an infinite direct sum and has infinite
dimension although each of its \irr\ subsectors has finite dimension.

Furthermore, the \irr\ \sps\ sectors are implemented by charged \itw s
$\psi_s$ as in Sect.\ 4, although eq.\ (4.7) is meaningless since
$\lam$ is infinite and the `dual' isometry $v$ does not exist.

It is therefore desirable to have a general theory for \qfl\ \noss\
of infinite index which establishes the existence of charged \itw s as
well as statements like Cor.\ 4.6., which do not refer to the index. On
the other hand, it is known that in the abstract theory of subfactors
there are counter examples \cite\Izpriv\ to these expectations, even
when there is a \cex. One needs therefore a criterium which `stabilizes'
the structure of the local subfactors (like compactness of the gauge
group) and which can be physically motivated.

Let us describe the expected structure which is known to hold with
compact gauge groups notwithstanding the index being infinite, along
with some problems to be overcome in the general infinite index case.
First, the commutant $\rho(N)' \cap N$ of the dual \cmor\ $\rho$ should
have the structure of an infinite direct sum of finite matrix rings
(corresponding to \irr\ subsectors $\rho_s$ with finite multiplicities),
and the minimal central projections should lie in the domain of the \ovw\
$\eta =j_N\comp\eps\inv\comp j_N \in P(N,N_1)$ (corresponding to finite
dimension of $\rho_s$). Furthermore, $\eta$ should be invariant under
$j_{N_1}$ which takes the projection corresponding to a subsector into the
projection corresponding to the conjugate subsector, in order to derive
the expected values $\eta(p_s) = d(\rho_s) = d(\bar\rho_s)$ on the
projections onto \irr\ subsectors. Apparently, the conjugation
invariance of $\eta$ does not hold {\it a priori} \cite\Izpriv. Second,
when the isometry $v$ as in Sect.\ 4 does not exist, a natural formula to
associate charged \itw s $\psi_s$ with observable \itw s $w_s \map
{\rho_s}\rho$ would be
$$ w_s = \g(\psi_s^*)w \qquad \Leftrightarrow \qquad \psi_s =
\g\inv\eta(ww_s^*). $$
If the index is finite, then these are equivalent to the anti-\imor\ (4.7)
(except for a normalization). For infinite index, however, $ww_s^*$ might
not be in the domain of $\eta$ even if $\eta(ww^*) = 1$ by definition,
cf.\ eq.\ (2.4), and $\eta(w_sw_s^*) = \eta(p_s) < \infty$ by assumption.

Note that if the existence of a charged \itw\ can be established
then it `implements' the \emor\ $\rho_s$ `in the average', i.e.,
$$ \rho_s(a) = \eps(\psi_s a \psi_s^*), $$
where the scalar $\eps(\psi_s\psi_s^*) \in \rho_s(\aa)'\cap\aa = \CC$
has been normalized to unity. This formula generalizes (4.8).

Let us tentatively call a subfactor which fulfils all the stated
expectations, a subfactor `of compact type'. The desired physical
criterium which compels a \qfl\ \nos\ to be `of compact type' is
expected to come from the split property \cite\DL\ for the net $\bb$
which is related to a `tame' field content and a decent thermodynamical
behaviour of the theory \cite\Nucl. It is well known that the stability
of the associated canonical intermediate type $I$ factor under gauge
\amor s implies that the gauge group must be compact \cite\DL, and
therefore the \nos\ associated with the gauge invariants is `of compact
type'. We may expect that a similar stability property holds for \cex s,
and that therefore a \qfl\ \nos\ with the split property is automatically
`of compact type' even in the absence of a gauge group.

The structure of subfactors with infinite index seems to be more stable
when the depth of the inclusion is 2. For a general study of this situation
(going along with a Kac-Hopf symmetry), see, e.g., \cite{\Hopf,\NW} and
references therein. In \cite\NW, an approximative substitute for the
dual isometry $v$ is constructed.
\newsec{Conclusions}
We have initiated a general theory of restriction and induction of \sps\
sectors between pairs of local quantum field theories $\aa\subset \bb$.
Here, the local \vna s $\aa(\oo)$ of `observables' in the space-time
region $\oo$ are specified as subalgebras
of the local \vna s $\bb(\oo)$ by a conditional expectation which is
consistent with restriction to subregions, and which preserves the
vacuum state. The latter conditions are considered as an abstraction
from the notion of an unbroken inner symmetry with respect to which the
observables $\aa$ are the invariants.

Our theory generalizes models with a compact gauge group \cite\DR\ and
confirms the results of previous specific model analysis (in the absence
of a gauge group) \cite{\Wass,\Coset}, which in fact anticipated the
general structures elaborated here. Our basic technical tool is the
\cmor\ associated with a subfactor, which provides an alternative approach
to the Jones extension more appropriate in the type $\III$ case.

Actually, the mathematical results are not restricted to the context of
quantum field theories. We have formulated them without reference to
locality properties, whenever the latter are irrelevant. In fact, our
theory hinges on Thm.\ 3.2.\ concerning a net with only two elements,
which plays a role as a `germ' for the really interesting nets.
Yet, due to the trivial \rep\ theory of the individual factors,
the results of Sect.\ 3 about \rep\ theory are relevant only if the nets
of \vna s do not possess a maximal element, but have a nontrivial
inductive limit (as in quantum field theory).

Among our results for the quantum field theoretical application is a
characterization of field algebras which extend (with finite index)
a given local theory of observables, in terms of observable quantities
only (Thm.\ 4.9.). We give examples for this characterization.
Furthermore, we relate the charged intertwiners in the field algebra
to basic quantities in the theory of subfactors, and derive operator
product expansions for the former (Prop.\ 4.7.).
\bgap\bigbreak
\baselineskip=14pt
{\bf Acknowledgements.} The authors have much benefitted from discussions
with and valuable criticism by K.\ Fredenhagen and J.E.\ Roberts, who are
both involved in a long-termed joint program on issues related to the
present analysis. K.-H.R.\ thanks the University of Rome II for hospitality
and the CNR for financial support during a visit in Rome. He is also
indebted to M.\ Izumi for several interesting discussions.
\bgap\bigbreak
{\bbf References}
\def\ref{\par \noindent \hangafter=1 \hangindent 15pt \cite}
\parskip 2pt
\baselineskip=2.5ex\ninepoint\smallskip
\def\it{\nineit}
\def\bf{\ninebf}
\def\CMP#1{Com\-mun.\ Math.\ Phys.\ {\bf #1}}

\def\IM#1{Invent.\ Math.\ {\bf #1}}
\def\JFA#1{J.\ Funct.\ Anal.\ {\bf #1}}
\def\JMP#1{Journ.\ Math.\ Phys.\ {\bf #1}}
\def\RMP#1{Rev.\ Math.\ Phys.\ {\bf #1}}
\ref\HK\ R.\ Haag, D.\ Kastler: {\it An algebraic approach to \qft},
\JMP{5}, 848--861 (1964).
\ref{\DR a} S.\ Doplicher, J.E.\ Roberts: {\it Fields, statistics and
non-abelian gauge groups}, \CMP{28}, 331--348 (1972).
\ref{\DR b} S.\ Doplicher, J.E.\ Roberts: {\it Why there is a field
algebra with a compact gauge group describing the \sps\ structure in
particle physics}, \CMP{131}, 51--107 (1990).
\ref\DHR\ S.\ Doplicher, R.\ Haag, J.E.\ Roberts: {\it Local
observables and particle statistics.\ I+II}, \CMP{23}, 199--230 (1971)
and {\bf 35}, 49--85 (1974).
\ref\Tak\ M.\ Takesaki: {\it Conditional expectations in \vna s},
\JFA{9}, 306--321 (1972).
\ref\BW\ J.J.\ Bisognano, E.H.\ Wichmann: {\it On the duality
condition for a hermitian scalar field}, \JMP{16}, 985--1007 (1975).
\ref\Wass\ A.\ Wassermann: {\it Subfactors arising from positive energy
\rep s of some infinite dimensional groups}, research notes (Cambridge 1990),
and {\it Operator algebras and conformal field theories}, research notes
(Cambridge 1994).
\ref{\FRS a} K.\ Fredenhagen, K.-H.\ Rehren, B.\ Schroer:
{\it Superselection sectors with braid group statistics and
exchange algebras.\ I}, \CMP{125}, 201--226 (1989).
\ref{\FRS b} K.\ Fredenhagen, K.-H.\ Rehren, B.\ Schroer:
{\it Superselection sectors with braid group statistics and
exchange algebras.\ II}, \RMP{Special Issue}, 113--157 (1992).
\ref\DL\ S.\ Doplicher, R.\ Longo: {\it Standard and split inclusions
of \vna s}, \IM{75}, 493--536 (1984).
\ref\CD\ F.\ Combes, C.\ Delaroche: {\it Groupe modulaire d'une
esp\'erance conditionelle dans une alg\`ebre de von Neumann},
Bull.\ Soc.\ Math.\ Franc.\ {\bf 103}, 385--426 (1975).
\ref\Hgp\ U.\ Haagerup: {\it Operator valued weights in \vna s. I+II},
\JFA{32}, 175--206 (1979) and {\bf 33}, 339--361 (1979).
\ref\Con\ A.\ Connes: {\it Spatial theory of \vna s}, \JFA{35}, 153--164
(1980).
\ref\Kos\ H.\ Kosaki: {\it Extension of Jones' theory on subfactors
to arbitrary factors}, \JFA{66}, 123--140 (1986).
\ref\Jon\ V.F.R.\ Jones: {\it Index for subfactors}, \IM{72}, 1--25 (1983).
\ref\PP\ M.\ Pimsner, S.\ Popa: {\it Entropy and index for
subfactors}, Ann.\ Sci.\ \'Ec.\ Norm.\ Sup.\ {\bf 19}, 57--106 (1986).
\ref\Hiai\ F.\ Hiai: {\it Minimizing indices of \cex s on a subfactor},
Publ.\ RIMS {\bf 24}, 673--678 (1988).
\ref{\Lon a} R.\ Longo: {\it Index of subfactors and statistics of
quantum fields.\ I}, \CMP{126}, 217--247 (1989).
\ref{\Lon b} R.\ Longo: {\it Index of subfactors and statistics of
quantum fields.\ II}, \CMP{130}, 285--309 (1990).
\ref\KL\ H.\ Kosaki, R.\ Longo: {\it A remark on the minimal index of
subfactors}, \JFA{107}, 458--470 (1992).
\ref\Hopf\ R.\ Longo: {\it A duality for Hopf algebras and for
subfactors.\ I}, \CMP{159}, 133--150 (1994).
\ref\Izu\ M.\ Izumi: {\it Application of fusion rules to the
classification of subfactors}, Publ.\ RIMS {\bf 27}, 953--994 (1991).
\ref\BDF\ D.\ Buchholz, C.\ D'Antoni, K.\ Fredenhagen: {\it The
universal structure of local algebras}, \CMP{111}, 123--135 (1987);
\newline J.\ Fr\"ohlich, F.\ Gabbiani: {\it Operator algebras and
conformal field theory}, \CMP{155}, 569--640 (1993);
\newline R.\ Brunetti, D.\ Guido, R.\ Longo: {\it Modular structure and
duality in conformal \qft}, \CMP{156}, 201--219 (1993).
\ref\Rob\ J.E.\ Roberts: {\it The statistical dimension, conjugation and
the Jones index}, contribution to this volume.
\ref\NCG\ A.\ Connes: ``Non-Commutative Geometry'', to be published
by Academic Press.
\ref\Popa\ S.\ Popa: {\it Correspondences}, preprint.
\ref\RS\ H.\ Reeh, S.\ Schlieder: {\it Bemerkungen zur
Unit\"ar\"aquivalenz von Lorentzinvarianten Feldern}, Nuovo Cim.\
(serie 10) {\bf 22}, 1051--1068 (1961).
\ref\BSM\ D.\ Buchholz, H.\ Schulz-Mirbach: {\it Haag duality in
conformal \qft}, \RMP{2}, 105--125 (1990).
\ref\Coset\ K.-H.\ Rehren: {\it Subfactors and coset models}, in:
``Generalized Symmetries in Physics'', H.-D.\ Doebner {\it et al.}\ eds.,
(World Scientific 1994), pp.\ 338--356.
\ref\RST\ K.-H.\ Rehren, Ya.S.\ Stanev, I.T.\ Todorov: {\it
Characterizing invariants for local extensions of current algebras},
preprint ESI 132 (1994) (Vienna) and DESY 94-164 (Hamburg).
\ref\Ocn\ A.\ Ocneanu: {\it Quantum symmetry, differential geometry of
finite graphs, and classification of subfactors}, Univ.\ Tokyo Seminary
Notes {\bf 45}, 1991 (notes by Y.\ Kawahigashi).
\ref\Spt\ K.-H.\ Rehren: {\it Space-time fields and exchange fields},
\CMP{132}, 461--483 (1990).
\ref\GL\ D.\ Guido, R.\ Longo: {\it Relativistic invariance and charge
conjugation in \qft}, \CMP{148}, 521--551 (1992); \newline
H.-W.\ Wiesbrock: {\it Conformal \qft\ and half-sided modular inclusions
of \vna s}, \CMP{158}, 537--544 (1993).
\ref\Izpriv\ M.\ Izumi: private communication.
\ref\Nucl\ D.\ Buchholz, E.H.\ Wichmann: {\it Causal independence and
the energy-level density of states in local \qft}, \CMP{106}, 321--344
(1986).
\ref\NW\ F.\ Nill, H.-W.\ Wiesbrock: {\it A comment on Jones inclusions
with infinite index}, contribution to this volume.
\bye